\documentclass{aa}  

\usepackage{mathtools}
\usepackage{mathrsfs}
\makeatletter
\newcommand{\ostar}{\mathbin{\mathpalette\make@circled\star}}
\newcommand{\make@circled}[2]{%
  \ooalign{$\m@th#1\smallbigcirc{#1}$\cr\hidewidth$\m@th#1#2$\hidewidth\cr}%
}
\newcommand{\smallbigcirc}[1]{%
  \vcenter{\hbox{\scalebox{0.77778}{$\m@th#1\bigcirc$}}}%
}
\usepackage{array}
\usepackage{makecell}
\newcolumntype{C}[1]{>{\centering\arraybackslash}p{#1}} % Centered column with cell aligned at the bottom
\newcolumntype{L}[1]{>{\raggedright\arraybackslash}m{#1}} % Left-flushed column with cell aligned at the middle
\newcolumntype{M}[1]{>{\centering\arraybackslash}m{#1}} % Centered column with text aligned at the middle
\newcolumntype{B}[1]{>{\centering\arraybackslash}b{#1}} % Centered column

\renewcommand{\refeq}[1]{Eq.~(\ref{#1})\xspace} % New definition from mathools...
\newcommand{\refeqs}[2]{Eqs.~(\ref{#1}) and (\ref{#2})\xspace}
\newcommand{\refeqfull}[1]{Equation~(\ref{#1})\xspace} % Full expansion of the name

\newcommand{\reffig}[1]{Fig.~\ref{#1}\xspace}
\newcommand{\reffigs}[2]{Figs.~\ref{#1} and~\ref{#2}\xspace} % 2 figures
\newcommand{\reffigfull}[1]{Figure~\ref{#1}\xspace} % Full expansion of the name
\newcommand{\reffigsfull}[2]{Figures~\ref{#1} and~\ref{#2}\xspace} % Full expansion of the name / 2 figures

\newcommand{\refsub}[1]{#1} % How to refere to a subfigure.
\newcommand{\refsubfig}[2]{Fig.~\ref{#1}\refsub{#2}\xspace}
\newcommand{\refsubfigfull}[2]{Figure~\ref{#1}\refsub{#2}\xspace} % Full expansion of the name
\newcommand{\refsubfigs}[2]{Figs.~\ref{#1}\refsub{#2}\xspace}
\newcommand{\refpan}[1]{Panel~#1} % How to refere to a panel in a figure caption.
\newcommand{\refpans}[1]{Panels~#1} % How to refere to a panel in a figure caption.

\newcommand{\reftab}[1]{Table~\ref{#1}\xspace}

\newcommand{\refsec}[1]{Sect.~\ref{#1}\xspace} % Section and subsection
\newcommand{\refsecs}[2]{Sects.~\ref{#1} and~\ref{#2}\xspace} % Section and subsection

\newcommand{\refapp}[1]{Appendix~\ref{#1}\xspace} % Appendix
\usepackage{graphicx}
\graphicspath{{./figures/}}

\usepackage{transparent}
\usepackage{xkeyval,xcolor}% http://ctan.org/pkg/{xkeyval,xcolor}
\makeatletter
\newlength{\sfp@hseplen}\newlength{\sfp@vseplen}
\define@cmdkey{subfigpos}[sfp@]{pos}[ul]{}% \sfp@pos
\define@cmdkey{subfigpos}[sfp@]{font}[\small]{}% \sfp@font
\define@cmdkey{subfigpos}[sfp@]{vsep}[0.75\baselineskip]{\setlength{\sfp@vseplen}{\sfp@vsep}}% \sfp@vsep
\define@cmdkey{subfigpos}[sfp@]{hsep}[3.5pt]{\setlength{\sfp@hseplen}{\sfp@hsep}}% \sfp@hsep
\newcommand{\subfigimg}[4][,]{%
        \setkeys{Gin,subfigpos}{pos,font,vsep,hsep,#1}% Set (default) keys
        \setbox1=\hbox{\includegraphics{#4}}% Store image in box
        \ifnum\pdfstrcmp{\sfp@pos}{ul}=0% UPPER LEFT placement of subfig label
                \leavevmode\rlap{\usebox1}% Print image
                \rlap{\hspace*{\sfp@hsep}\raisebox{\dimexpr\ht1-\sfp@vsep}{\transparent{#3}{\setlength{\fboxsep}{1pt}\colorbox{white}{%
\transparent{1}\sfp@font{#2}}}%
}}% Print label
                \phantom{\usebox1}% Insert appropriate spacing
        \else\ifnum\pdfstrcmp{\sfp@pos}{ur}=0% UPPER RIGHT placement of subfig label
                \leavevmode\usebox1% Print image
                \llap{\raisebox{\dimexpr\ht1-\sfp@vsep}{\sfp@font{#2}}\hspace*{\sfp@hsep}}% Print label
        \else\ifnum\pdfstrcmp{\sfp@pos}{lr}=0% LOWER RIGHT placement of subfig label
                \leavevmode\usebox1% Print image
                \llap{\raisebox{\sfp@vsep}{\sfp@font{#2}}\hspace*{\sfp@hsep}}% Print label
        \else% Assume LOWER LEFT placement of subfig label
                \leavevmode\rlap{\usebox1}% Print image
                \rlap{\hspace*{\sfp@hseplen}\raisebox{\sfp@vsep}{\sfp@font{#2}}}% Print label
                \phantom{\usebox1}% Insert appropriate spacing
        \fi\fi\fi
}
\newcommand{\fontfig}[1]{\tiny$\!\!$\color{#1}\textbf}
\newcommand{\AspectRatio}[1]{\dimexpr 1pt * \wd#1 / \ht#1 \relax} % Aspect ratio of a subfigure
\DeclarePairedDelimiterX{\paren}[1]{(}{)}{#1}
\newcommand{\Paren}[1]{\paren*{#1}}
\let\brace=\undefined % Redifine \brace
\DeclarePairedDelimiterX{\brace}[1]{\{}{\}}{#1}
\newcommand{\Brace}[1]{\brace*{#1}}
\let\brack=\undefined % Redifine \brack
\DeclarePairedDelimiterX{\brack}[1]{[}{]}{#1}
\newcommand{\Brack}[1]{\brack*{#1}}
\DeclarePairedDelimiterX{\bbrack}[1]{\llbracket}{\rrbracket}{#1}%  [| ... |]

\DeclarePairedDelimiterX{\abs}[1]{\rvert}{\lvert}{#1}     %  | ... |
\newcommand{\Abs}[1]{\abs*{#1}}
\DeclarePairedDelimiterX{\norm}[1]{\lVert}{\rVert}{#1}    % || ... ||

\DeclarePairedDelimiterX{\avg}[1]{\langle}{\rangle}{#1}   %  < ... >
\newcommand{\Avg}[1]{\avg*{#1}}

\DeclarePairedDelimiterX{\ceil}[1]{\lceil}{\rceil}{#1}     % ceil operator

\DeclarePairedDelimiterX{\floor}[1]{\lfloor}{\rfloor}{#1}  % floor operator

\usepackage{algorithm}
\usepackage[noend]{algpseudocode}
\newcommand{\crosscorr}{\ostar}                         % Correlation symbol

\newcommand{\imag}[1]{\mathscr{I}\Brack{#1}} % Imaginary part
\newcommand{\real}[1]{\mathscr{R}\Brack{#1}} % Real part
\newcommand{\Arg}[1]{\mathscr{A}\Brack{#1}}  % Argument

\newcommand{\Tag}[1]{\text{#1}}                 % Tag
\newcommand{\V}[1]{{\boldsymbol{#1}}}   % vector (with amsmath)
\newcommand{\Inv}{^{-1}}                % inverse
\newcommand{\conj}[1]{\overline{#1}}    % conjugate
\newcommand{\scaprod}{^T\!\!\cdot}            % scalar product

\newcommand{\x}{{x}}
\newcommand{\Vx}{{\V{\x}}}
\newcommand{\Vk}{{\V{k}}} 
\newcommand{\Vv}{{\V{v}}} 
\newcommand{\Vdelta}{{\V{\delta}}} 

\newcommand{\Vm}{\V{m}}
\newcommand{\Vw}{\V{w}}
\newcommand{\Vn}{\V{n}}
\newcommand{\Vc}{\V{c}}
\newcommand{\Vp}{\V{p}}

\newcommand{\muconj}{\conj{\mu}}
\newcommand{\muf}{\fdep{\mu}}
\newcommand{\mufconj}{\conj{\muf}}
\newcommand{\Deltaf}{\fdep{\Delta}}

\newcommand{\fdep}[1]{\text{\b{$#1$}}}
\newcommand{\mf}[1]{\fdep{m}_{#1}}
\newcommand{\mfconj}[1]{\conj{\fdep{m}}_{#1}}
\newcommand{\wf}[1]{\fdep{w}_{#1}}
\newcommand{\nf}[1]{\fdep{n}_{#1}}
\newcommand{\nfconj}[1]{\conj{\fdep{n}}_{#1}}
\newcommand{\cf}[1]{\fdep{c}_{#1}}

\newcommand{\pf}[1]{\fdep{p}_{#1}}
\newcommand{\pfconj}[1]{\conj{\fdep{p}}_{#1}}

\newcommand{\Vvar}[1]{\V{\mathscr{V}}_{\!\!\!#1}} 
\newcommand{\corr}[1]{\mathscr{C}_{\!#1}} 
\newcommand{\Vcorr}[1]{\V{\mathscr{C}}_{\!#1}} 
\newcommand{\corrt}[1]{\tilde{\mathscr{C}}^\Tag{#1}} 
\newcommand{\Vcorrt}[1]{\tilde{\V{\mathscr{C}}}^\Tag{#1}} 
\newcommand{\argmin}[2]{\underset{#1}{\text{argmin}}\;#2}
\newcommand{\argmax}[2]{\underset{#1}{\text{argmax}}\;#2}
\newcommand{\dirfunc}[1]{\perp_{#1}}
\newcommand{\TF}[1]{\mathscr{F}_{\!#1}}

\newcommand{\ct}{\cos\theta}
\newcommand{\st}{\sin\theta}

\newcommand{\CM}{\V{M}^\Tag{cmd}}
\newcommand{\IM}[1]{\V{M}^\Tag{int}_{#1}}
\newcommand{\IMtilde}[1]{\tilde{\V{M}}^\Tag{int}_{#1}}

\newcommand{\rFried}{r_{0}}
\newcommand{\vWind}{v_{0}}
\newcommand{\thetaWind}{\theta_{0}}
\newcommand{\mKL}{m^{\Tag{KL}}}
\newcommand{\DMpitch}{\Delta^{\Tag{DM}}}
\newcommand{\SApitch}{\Delta}
\newcommand{\nmodes}{n^\Tag{mod}}
\newcommand{\nact}{n^\Tag{act}}
\newcommand{\dact}{d^\Tag{act}}
\newcommand{\kctrl}{\mathcal{K}^\Tag{ctrl}}
\newcommand{\fwind}{f^\parallel}

\newcommand{\numKL}{35}
\newcommand{\numKLbench}{40}
\newcommand{\thetaGain}{60}

\newcommand{\weight}{{w}}
\newcommand{\weightWFS}{\weight^{\Tag{WFS}}}
\newcommand{\weightvalid}{\weight^{\Tag{valid}}}

\newcommand{\mUS}{m^{\Tag{up}}}

\newcommand{\tDM}{\tau^{\Tag{DM}}}
\newcommand{\tlat}{\tau^{\Tag{lat}}}
\newcommand{\tRTC}{\tau^{\Tag{RTC}}}
\newcommand{\tWFS}{\tau^{\Tag{WFS}}}
\newcommand{\gint}{g^{\int}}
\newcommand{\gleak}{g^{\epsilon}}
\newcommand{\gcl}{g^{\Tag{cl}}}

\newcommand{\nph}{n^{\Tag{ph}}}

\newcommand{\X}[1]{$#1\!\times\!#1$}

\usepackage{siunitx}
\newcommand{\percent}[1]{#1\,\%}     % percent
\newcommand{\Nguu}{\includegraphics[height={\f@size pt*2/3}]{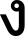}}
\newcommand{\Reua}{\includegraphics[height={\f@size pt*2/3}]{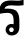}}
\newcommand{\Thong}{\includegraphics[height={\f@size pt*2/3}]{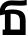}}
\usepackage{txfonts}
\usepackage[colorlinks=true, allcolors=blue]{hyperref}
\usepackage{flushend}
\usepackage{silence}
\begin{document}

%-------------------------------------------------------------------%
%------------------------------ Title ------------------------------%
%-------------------------------------------------------------------%

\title{Estimation of the lateral mis-registrations of the \\ GRAVITY+ adaptive optics system}
\subtitle{Perturbative method with open-loop modal correlation and \\ non-pertubative method with temporal correlation of closed-loop telemetry}

\author{
    A.~Berdeu\inst{\ref{lesia}}
    \and H.~Bonnet\inst{\ref{esog}}
    \and J.-B.~Le~Bouquin\inst{\ref{ipag}}
    \and C.~Édouard\inst{\ref{lesia}}
    \and T.~Gomes\inst{\ref{centra},\ref{porto}}
    \and P.~Shchekaturov\inst{\ref{esog}}
    \and R.~Dembet\inst{\ref{lesia}}
    \and T.~Paumard\inst{\ref{lesia}}
    \and S.~Oberti\inst{\ref{esog}}
    \and J.~Kolb\inst{\ref{esog}}
    \and F.~Millour\inst{\ref{cotedazur}}
    \and P.~Berio\inst{\ref{cotedazur}}
    \and O.~Lai\inst{\ref{cotedazur}}
    \and F.~Eisenhauer\inst{\ref{mpe}}
    \and P.~Garcia\inst{\ref{centra},\ref{porto}}
    \and C.~Straubmeier\inst{\ref{cologne}}
    \and L.~Kreidberg\inst{\ref{mpia}}
    \and S.~F.~Hönig\inst{\ref{southamp}}
    \and D.~Defrère\inst{\ref{kuleuven}}
}

\institute{
    LESIA, Observatoire de Paris, Université PSL, Sorbonne Université, Université Paris Cité, CNRS, 5 place Jules Janssen, 92195 Meudon, France 
    \label{lesia}
    \and
    European Southern Observatory, Karl-Schwarzschild-Straße 2, 85748 Garching, Germany
    \label{esog}
    \and
    Univ. Grenoble Alpes, CNRS, IPAG, 38000 Grenoble, France
    \label{ipag}
    \and
    CENTRA - Centro de Astrof\' isica e Gravita\c c\~ao, IST, Universidade de Lisboa, 1049-001 Lisboa, Portugal
    \label{centra}
    \and
    Universidade do Porto, Faculdade de Engenharia, Rua Dr.~Roberto Frias, 4200-465 Porto, Portugal
    \label{porto}
    \and
    Université Côte d’Azur, Observatoire de la Côte d’Azur, CNRS, Laboratoire Lagrange, France
    \label{cotedazur}
    \and
    Max Planck Institute for extraterrestrial Physics, Giessenbachstra\ss e~1, 85748 Garching, Germany
    \label{mpe}
    \and
    $1^\Tag{st}$ Institute of Physics, University of Cologne, Z\"ulpicher Stra\ss e 77, 50937 Cologne, Germany
    \label{cologne}
    \and
    Max Planck Institute for Astronomy, K\"onigstuhl 17, 69117 Heidelberg, Germany
    \label{mpia}
    \and
    School of Physics \& Astronomy, University of Southampton, University Road, Southampton SO17 1BJ, UK
    \label{southamp}
    \and
    Institute of Astronomy, KU Leuven, Celestijnenlaan 200D, B-3001, Leuven, Belgium
    \label{kuleuven}
    \\
    \email{anthony.berdeu@obspm.fr}
}

\date{Received at some point in 2024 / Accepted hopefully soon afterwards}

% \abstract{}{}{}{}{} 
% 5 {} token are mandatory
 
\abstract
{
The GRAVITY+ upgrade implies a complete renewal of its adaptive optics (AO) systems. Its complex design, featuring moving components between the deformable mirrors and the wavefront sensors, requires the monitoring and auto-calibrating of the lateral mis-registrations of the system while in operation.
}
{
For preset and target acquisition, large lateral registration errors must be assessed in open loop to bring the system to a state where the AO loop closes. In closed loop, these errors must be monitored and corrected, without impacting the science.
}
{
With respect to the first requirement, our method is perturbative, with two-dimensional modes intentionally applied to the system and  correlated to a reference interaction matrix. For the second requirement, we applied a non-perturbative approach that searches for specific patterns in temporal correlations in the closed loop telemetry. This signal is produced by the noise propagation through the AO loop.
}
{
Our methods were validated through simulations and on the GRAVITY+ development bench. The first method robustly estimates the lateral mis-registrations, in a single fit and with a sub-subaperture resolution while in an open loop. The second method is not absolute, but it does successfully bring the system towards a negligible mis-registration error, with a limited turbulence bias. Both methods proved to robustly work on a system still under development and not fully characterised. 
}
{
Tested with Shack-Hartmann wavefront sensors, the proposed methods are versatile and easily adaptable to other AO instruments, such as the pyramid, which stands as a baseline for all future AO systems. The non-perturbative method, not relying on an interaction matrix model and being sparse in the Fourier domain, is particularly suitable to the next generation of AO systems for extremely large telescopes that will present an unprecedented level of complexity and numbers of actuators.
}


%\titlerunning{}

\maketitle

%-------------------------------------------------------------------%
%----------------------------- Article -----------------------------%
%-------------------------------------------------------------------%

% Removing shyperref warnings "Suppressing link with empty target" by making all '\pageref's default to hyperlink-less references
\let\oldpageref\pageref
\renewcommand{\pageref}{\oldpageref*}

\section{Introduction}
\label{sec:intro}

GRAVITY+ \citep{GRAVITYplus:22_messenger} is a combined upgrade of the GRAVITY instrument and of its host observatory the Very Large Telescope Interferometer \citep[VLTI,][]{GRAVITY:17_VLTI} of the European Southern Observatory (ESO). Among other work packages, it features a major update of its adaptive optics (AO) systems, with a renewal of all the \mbox{instances} installed on each of the four unit telescopes (UTs) of the VLTI.

The role of an AO system is to compensate in real time for the wavefront aberrations induced by atmospheric turbulence \citep{Roddier:99_AO_system}. These phase aberrations are measured by a wavefront sensor (WFS) whose signals are analysed by a real time computer (RTC) and turned into a set of commands sent to a deformable mirror (DM) that compensates the optical aberrations. To be effective, this feedback loop must run faster than the turbulence temporal evolution, with typical frequencies ranging from several hundred Hertz to a kilo-Hertz.

In the case of the GRAVITY+ adaptive optics \citep[GPAO,][]{LeBouquin:23_GPAO_design}, the RTC is based on the upgrade of the Standard Platform for Adaptive optics Real Time Applications \cite[SPARTA-upgrade,][]{Suarez:12_SPIE_SPARTA, Shchekaturov:23_SPARTA_upgrade, Dembet:23_RTC_GPAO} which inherits the high level functionalities developed for the Spectro Polarimetric High-contrast Exoplanet REsearcher \citep[SPHERE,][]{Beuzit:19_SPHERE} and the Adaptive \mbox{Optics} Facility \citep[AOF,][]{Oberti:18_SPIE_AO_in_AOF}. Overall, GPAO features a \X{43} DM with about 1200 actuators within the \SI{100}{\milli\meter} pupil and Shack-Hartmann wavefront sensors \citep[SH-WFSs,][]{Shack:71_SHWFS}: a \X{30} laser guide star (LGS) WFS, and a vari-ety of natural guide stars (NGS) WFSs (visible \X{40}, visible \X{4}, infrared \X{9}). The pupil of the system is given by the secondary mirror (M2) of the telescope, defining the beam size and position%, and located in the elevation gondola \SOB{elevation gondola: never heard of this term} of the telescope
. The DM is located in the Coudé train of the telescope, rotating with the azimuth, while the WFSs are at the Coudé focus, fixed to the ground. Moreover, a number of sub-system within the AO WFS have to also track the azimuth and/or the elevation angles (a K-mirror, an atmospheric dispersion compensator, and the table used to patrol the field). The elevation and the azimuth rotations of the telescope itself have wobble amplitude of 2\% of the pupil diameter. The tracking sub-systems within the WFS have typical wobble amplitude of 5\% of the pupil diameter. As a consequence of this configuration, for the most resolving SH-WFS (\X{40}), the combined wobble amplitude corresponds to several subapertures.

This significant wobbling of the system leads to two important requirements for GPAO: (I)~GPAO needs a means to quickly estimate large lateral-registration during the acquisition of the target to within an accuracy to achieve stable closed loop operation, controlling most of the observable modes \citep[typically \percent{20} of the subaperture size, see for instance][]{Oberti:16_SPIE_AOF_LTAO}; (II)~during the subsequent observation, the azimuth and elevation angle continues to evolve, and so do their intrinsic wobbles, and GPAO needs a mean to track this small and continuously changing lateral-registration, while the AO loop is closed. 
% Consequently, (i) active control of the DM/WFS lateral-registration at a level of \percent{20} of the sub-aperture diameter is necessary to ensure loop stability at instrument preset and through operation; and (ii) since the DM is \emph{not} the instrument pupil, the imprint of the photometric pupil in the WFS cannot be used to track this lateral-registration. This is also mandatory to be able to operate the system on the internal light source of the VLTI which does not properly define a photometric pupil because it is located at the Nasmyth focus (that is after M2).
However, because the DM is  not the instrument pupil, the imprint of the photometric pupil in the WFS cannot be used to track this lateral-registration, as is typically done in most AO systems. This is even more applicable when the system operates on the internal light source of the VLTI, which does not properly define a photometric pupil because of its location at the Nasmyth focus (i.e. after M2). Therefore, there is a need for a proper, direct measurement of the DM/WFS lateral-registration at a level of \percent{20} of the sub-aperture diameter at instrument preset and through operation.

One complete instance of GPAO out of the four is currently being integrated and tested at the Lagrange laboratory in Nice \citep{Millour:22_SPIE_GPAO_bench}. The test bench features multiple sources, a turbulent phase screen, a pupil mask mimicking the VLT, and a beam expander to properly illuminate the DM and the SH-WFSs with the correct beam size and f-number. The bench permits \mbox{operation} of GPAO in both NGS and LGS modes, with the caveat that the LGS source does not incorporate spot elongation and cone effect. The results presented in this paper focus on the NGS VIS WFS (\X{40}, point source) mode but all tests have also been performed with the LGS WFS as well (\X{30}, extended source).

Regarding the mis-registration errors, the registration of a couple DM/WFS is defined by how the actuator geometry of the former is seen by the optics geometry of the latter\footnote{For example the Fried geometry~\citep{Fried:77}, where the DM actuators are optically placed at the corner of the SH-WFS subapertures.}. This global geometry is defined in the design phase of the instrument. A mis-registration error occurs when there is a mismatch between the designed values and the real system. Among others \citep[Sect. 2.1]{Heritier:19_PHD}, the classical (and most impacting) mis-registrations are the $x$ and $y$-shifts (lateral translation of the DM with respect to the WFS), the clocking (rotation of the DM with respect to the WFS), and the magnification and \mbox{anamorphosis} (stretches of the DM with respect to the WFS).

Different techniques exist to assess the mis-registration state of an AO system. For instance, in the case where the DM is also the pupil of the instrument, the WFS photometry can be used to track the lateral error \citep{Kolb:16_SPIE_Review_AO_calibration}. However, most of the tools to track mis-registrations use the interaction matrix (IM) of the AO system \citep{Roddier:99_AO_system}. Close to its functioning point, an AO system is assumed to be linear and the IM links the \mbox{commands} sent to the DM to the measures provided by the WFS\footnote{Or possibly multiple DMs and multiple WFSs depending on the considered system.}. % Classically, AO systems are defined by their interaction matrices (IMs) \cite{Roddier:99_AO_system}. Closed to their functioning point, AO systems are assumed to be linear and IMs link the commands sent to the DMs to the measures provided by the WFSs.
If the AO system is sensitive to the considered mis-registration, an alignment error should then leave a signature in its IM.

In early AO systems, with a limited number of actuators and with an internal calibration source providing high signal-to-noise ratios (S/Ns), IMs were experimentally measured and the AO systems were run with `as-built', rather than `as-designed' IMs \citep{Oberti:06_SPIE_PSIM_vs_measured_IM}. Doing so, any mis-registration error is calibrated and taken into account by the AO loop. The arrival of telescopes with adaptive secondary mirrors, such as the Multiple Mirror Telescope \citep{Brusa:03_MMT}, the Large Binocular Telescope \citep{Riccardi:03_SPIE_LBT_SDM}, or the AOF \citep{Strobele:06_SPIE_AOF}, complicated the identification of registration errors. Indeed, without any focal plane available upstream of the DM, and thus no reference source, the AO system must be calibrated directly on-sky at night. Different techniques to measure the full IM were implemented, such as fast `push-pull' commands to freeze the turbulence or command modulation and signal demodulation to increase the S/N \citep{Oberti:06_SPIE_PSIM_vs_measured_IM, Pinna:12_SPIE_FLAO, Lai:21_DO_CRIME}. But such IMs are generally noisy and depend on the experimental conditions \citep{Kolb:16_SPIE_Review_AO_calibration}. In addition, night time is precious and the increase of AO system complexity is synonymous to longer and more sensitive calibrations.

To tackle these issues, new methods emerged, based on the fact that a physical modelling of the interaction matrix is generally available, so-called synthetic IM \citep{Oberti:06_SPIE_PSIM_vs_measured_IM, Heritier:18_PyWFS_calibration}. By definition, these matrices are noiseless but sensitive to model errors, that is to say a mismatch between the numerical model and the real system \citep{Kolb:12_SPIE_AOF}. If the first order of WFS physics is correctly implemented, the mis-registrations become this main source of error \citep{Heritier:19_PHD} and the synthetic model thus only depends on a limited number of parameters. These parameters can be fitted through calibrations to obtain pseudo-synthetic interaction matrices (PSIMs), where the `pseudo' emphasises that inputs from the real system are used in the synthetic model \citep{Oberti:06_SPIE_PSIM_vs_measured_IM}.

In general, the parameter fitting is performed on IMs acquired for this purpose during dedicated calibration procedures, on internal sources or on-sky %\SOB{I believe that the first time that the IM was used to estimate registration was on MACAO-VLTI anb SINFONI based on an idea from Henri. See Oberti 2004: Calibration of a curvature sensor/bimorph mirror AO system: interaction matrix measurement on MACAO systems} 
\citep{Oberti:04_SPIE_IM_misreg, Neichel:12_SPIE_MCAO, Kolb:16_SPIE_Review_AO_calibration, Heritier:18_PyWFS_calibration}. Nonetheless, mis-registrations are susceptible to evolve during observation due to mechanical flexion or thermal evolution. And future Extremely Large Telescopes \citep[ELTs,][]{Johns:06_SPIE_GMT, Gilmozzi:07_ELT, Boyer:18_SPIE_TMT} will bring this challenge to another level, with unprecedented distances between DMs and WFSs, with potentially different moving/rotating parts in-between, prone to misalignment. It then becomes impossible to perform regular calibrations and it is thus necessary to track the evolution of the mis-registrations directly during the scientific acquisitions with online tools for AO system auto-calibration. A strategy that has proven to be effective is recovering the IM directly from AO telemetry data \citep{Bechet:12_SPIE, Kolb:12_SPIE_AOF}. These methods introduced the fact that the measurement noise propagates through the AO loop, producing meaningful signal from which the IM can be estimated. Nonetheless, a large amount of telemetry data must be gathered so that the IM structure emerges from the loop noise, strongly increasing the recording and computation times. In addition, \citet[Sect. 4.3]{Heritier:19_PHD} showed that such an approach can be corrupted by the temporal error of the AO loop, with a bias induced by a frozen flow turbulence.

To further save time while increasing the S/N and debiasing the error estimation, recent works have started to focus on partial IM acquisition, focussing on dedicated and well chosen modes injected in the AO loop \citep{Heritier:21_SPRINT}. Contrary to online IM estimation, this method is invasive and may corrupt the science depending on the chosen modes and the amplitude to apply to get meaningful S/N. A trade-off must be balanced, for example, by applying the disturbance during the shutter closing time of the science instrument if possible. Such an invasive method is the current baseline of the AO systems of the first generation of instruments of the ESO ELT.

Finally, all these methods are based on the minimisation of the difference between the measured IM and the PSIM. Such PSIM models are generally complex and not directly invertible. Under the assumption that the mis-registration errors are small, the PSIMs are linearised close to their functioning point to get sensitivity matrices, leading to an iterative fit of the parameters, potentially laborious and slow and with a limited capture range \citep{Kolb:12_SPIE_AOF, Neichel:12_SPIE_MCAO, Heritier:18_PyWFS_calibration}. There have been several variants or improvements of this technique in recent years. \citet[Sect. 2.5.4]{Heritier:19_PHD} recomputes iteratively the sensitivity matrix in order to increase the capture range and achieve a larger linearity. \citet{Heritier:21_SPRINT} performed a numerical demonstration that such modal IMs, and thus the mis-registrations, can be estimated in a closed loop with a dedicated perturbation in the command, while controlling the linearity sensitivity of the estimator; however the applicability  of this method in the lab or on-sky is yet to be confirmed.

To meet the two requirements of GPAO introduced above: (1)~there is still a lack of a fast and robust method with a large capture range and that works in open loop to quickly align the system at preset and during science target acquisition; and (2)~during operation, we still miss a fast\footnote{We notice that this constraint on the speed could be relaxed for the VLTI where the physics of the mis-registrations is bound to be slow after the preset. Nonetheless, we aim to test on-sky the method efficiency in preparation of the future ELTs where the timescale of the misalignment evolution will be much shorter.} and non-invasive tool to monitor the lateral error, with the additional constraints that the solutions must be adaptable to the SPARTA architecture. To answer these two requirements, we present (in {\refsecs{sec:main_2D_corr}{sec:main_CL_estim}} of this paper) two new methods to fit the lateral errors in the system. The first one is based on a perturbative approach, whereby two-dimensional (2D) modes are applied on the system in open loop in order to be spatially correlated to a reference. The second one is a non-perturbative approach working in closed loop by analysing correlation signals in the telemetry of the commands.

We present the methods in \refsecs{sec:method_2D_corr}{sec:method_CL_estim}. For each method,  (i)~its general idea is given; (ii)~then its theoretical steps are developed; and (iii) finally, we summarise its main advantages. Then, in \refsecs{sec:simu_2D_corr}{sec:simu_CL_estim}, we assess the precision and the limits of the methods via end-to-end simulations. Finally, in \refsecs{sec:exp_2D_corr}{sec:exp_CL_estim}, we validate our approaches on a real system in the GPAO development bench.

\section{Perturbative 2D modal estimator}
\label{sec:main_2D_corr}

\subsection{Proposed method}
\label{sec:method_2D_corr}

\subsubsection{General concept}

\begin{figure}[t!] % fig:2D_corr_KL
        \centering
        
        % Internal command of the figure for the automatic sizing
        % Path of the files
        \newcommand{\PathFig}{Fig_2D_corr_KL/}
        
        % Line ratio
        \newcommand{\LineRatio}{0.975}
        
        % Font of the text in the figure
        \newcommand{\fontTxt}[1]{\textbf{\tiny #1}}
        
        % Width of the text boxes        
        \newcommand{\widthTxt}{8pt}
        
        % Vertical space between lines
        \newcommand{\spaceLine}{-0.1cm}
        
        \newcommand{\widthFig}{\dimexpr (\linewidth - \widthTxt)}
        
        \newcommand{\subfigColor}{white}
        
        % Suffix KL
        \newcommand{\numKLone}{6}
        \newcommand{\numKLtwo}{20}
        \newcommand{\numKLthree}{35}
        \newcommand{\numKLfour}{46}

        % Getting the size of the boxes
        \sbox1{\includegraphics{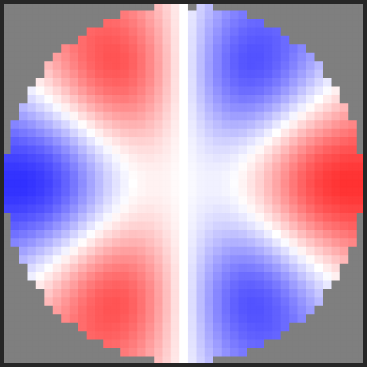}}
        \sbox2{\includegraphics{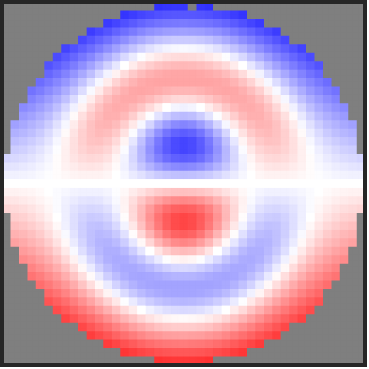}}
        \sbox3{\includegraphics{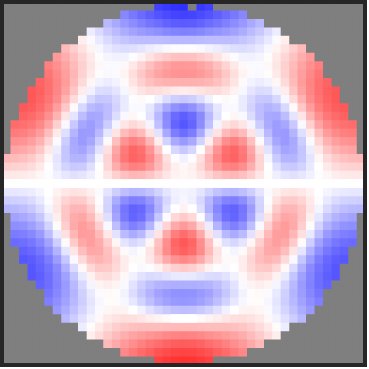}}
        \sbox4{\includegraphics{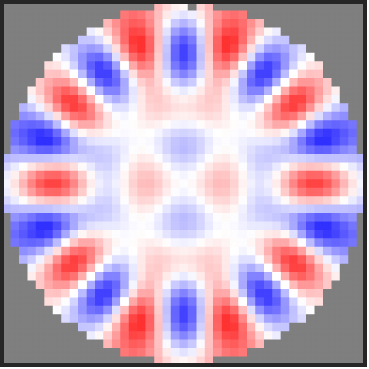}}
        
        % Defining column width command
        \newcommand{\ColumnWidth}[1]
                {\dimexpr \LineRatio \widthFig * \AspectRatio{#1} / (\AspectRatio{1} + \AspectRatio{2} + \AspectRatio{3} + \AspectRatio{4}) \relax
                }
        \newcommand{\ColumnGap}{\hspace {\dimexpr \widthFig /4 - \LineRatio\widthFig /4 }}

        % Figure table
        \begin{tabular}{
                @{}
                M{\widthTxt}
                @{\ColumnGap}
                M{\ColumnWidth{1}}
                @{\ColumnGap}
                M{\ColumnWidth{2}}
                @{\ColumnGap}
                M{\ColumnWidth{3}}
                @{\ColumnGap}
                M{\ColumnWidth{4}}
                @{}
                }
                
                % Title line
                &
                \fontTxt{$\mKL=\numKLone$}
                &
                \fontTxt{$\mKL=\numKLtwo$}
                &
                \fontTxt{$\mKL=\numKLthree$}
                &
                \fontTxt{$\mKL=\numKLfour$}
                \\                
                
                % DM command
                \rotatebox[origin=l]{90}{\fontTxt{DM command}} &
                \subfigimg[width=\linewidth,pos=ul,font=\fontfig{\subfigColor}]{}{0.0}{\PathFig DM_KL_\numKLone} &
                \subfigimg[width=\linewidth,pos=ul,font=\fontfig{\subfigColor}]{}{0.0}{\PathFig DM_KL_\numKLtwo} &
                \subfigimg[width=\linewidth,pos=ul,font=\fontfig{\subfigColor}]{}{0.0}{\PathFig DM_KL_\numKLthree} &
                \subfigimg[width=\linewidth,pos=ul,font=\fontfig{\subfigColor}]{}{0.0}{\PathFig DM_KL_\numKLfour}
                \\[\spaceLine]             
                
                % x-slopes
                \rotatebox[origin=l]{90}{\fontTxt{WFS $y$-slopes}} &
                \subfigimg[width=\linewidth,pos=ul,font=\fontfig{\subfigColor}]{}{0.0}{\PathFig WFS_KL_\numKLone} &
                \subfigimg[width=\linewidth,pos=ul,font=\fontfig{\subfigColor}]{}{0.0}{\PathFig WFS_KL_\numKLtwo} &
                \subfigimg[width=\linewidth,pos=ul,font=\fontfig{\subfigColor}]{}{0.0}{\PathFig WFS_KL_\numKLthree} &
                \subfigimg[width=\linewidth,pos=ul,font=\fontfig{\subfigColor}]{}{0.0}{\PathFig WFS_KL_\numKLfour}
                \\[\spaceLine]         
                
                % Auto-correlation
                \rotatebox[origin=l]{90}{\fontTxt{Auto-correlation}} &
                \subfigimg[width=\linewidth,pos=ul,font=\fontfig{\subfigColor}]{}{0.0}{\PathFig 2D_corr_KL_\numKLone} &
                \subfigimg[width=\linewidth,pos=ul,font=\fontfig{\subfigColor}]{}{0.0}{\PathFig 2D_corr_KL_\numKLtwo} &
                \subfigimg[width=\linewidth,pos=ul,font=\fontfig{\subfigColor}]{}{0.0}{\PathFig 2D_corr_KL_\numKLthree} &
                \subfigimg[width=\linewidth,pos=ul,font=\fontfig{\subfigColor}]{}{0.0}{\PathFig 2D_corr_KL_\numKLfour}
                \\
                
        \end{tabular}
              
        \caption{\label{fig:2D_corr_KL} General idea of the spatial 2D correlation method for different KL modes $\mKL\in\Brace{\numKLone, \numKLtwo, \numKLthree, \numKLfour}$. When different modes are applied on the DM (\textit{first line}), the patterns seen by the WFS show different spatial scales (\textit{second line}). The grey pixels correspond to the missing slopes~$\weightWFS\Paren{\Vx}=0$ in the PSIM model that are hidden by the pupil. A lateral mis-registration would shift the peak in the correlation between the PSIM and the measured IM (\textit{third line}).}
\end{figure}

A SH-WFS produces spatial measurements of the input wavefront's gradient. Thus, when a spatial pattern is applied on the DM, a specific spatial pattern will be seen in the SH-WFS data. The pattern applied on the DM can be its individual influence functions (zonal IM) or dedicated modes such as Zernike polynomials or Karhunen–Loève (KL) functions~\citep{Dai:95_KL} obtained by combining DM commands (modal IM). \reffigfull{fig:2D_corr_KL} presents some KL modes applied in the DM space (first line) and their counter-part on the $x$-slopes of the SH-WFS (second line) when reshaped in 2D. As expected, the patterns in the 2D representation of SH-WFS slopes are highly structured with a pattern specific to each mode.

The idea of the proposed method is to use IMs that are Nyquist-sampled, in the sense that the highest spatial frequency in the generated gradients is sampled by at least two subapertures. In this case, a lateral mis-registration in the system can be approximated by a, possibly sub-subaperture, geometric shift of the measurements. Thus, the 2D spatial correlation of a measured modal IM with a reference modal IM should be usable as a lateral error estimator. Contrary to other solutions based on modal perturbations, such as the one proposed by \citet{Heritier:21_SPRINT}, this method does not use a synthetic IM model fitting that implies the tuning of its parameters, but it uses the 2D representation of the measured IMs to directly estimate the parameters of interest.

Nonetheless, as seen in the auto-correlation of the KL modes in \reffig{fig:2D_corr_KL} (third line), the sensitivity of the correlation to a misalignment will depend on the considered mode, as already mentioned by \citet{Heritier:21_SPRINT}. Low spatial frequency modes, such as $\mKL=6$, produce a single but extended correlation peak while high spatial frequency modes, such as $\mKL=35$, produce localised but multiple correlation peaks. By working with different modes, our method aims at combining the large capture range of the low spatial frequencies with the sensitivity of the high spatial frequencies. The only constrain is to use KL modes whose gradients are properly sampled by the WFS.

\subsubsection{Estimation of the registration error from a modal IM}

As explained in the previous section, the method implies to reshape the slopes of the modal IM on 2D spatial positions $\Vx=\Paren{x,y}$ lying on a 2D Cartesian grid of pitch $\SApitch$, the subaperture size. Nonetheless, on such a square grid, some slopes are missing in the model (pupil external edge and central obscuration by the secondary mirror) and are set to zero. They are highlighted in grey in \reffig{fig:2D_corr_KL}. In the following, we note this mask of modelled slopes:
\begin{equation}
        \label{eq:weight_WFS}
        \weightWFS\Paren{\Vx} = 
        \begin{cases}
                1 \text{ if the slope at~$\Vx$ is in the IM,}
                \\
                0 \text{ otherwise.}
        \end{cases}
\end{equation}
On top of the slopes outside the pupil, partially illuminated subapertures of the SH-WFS, noisier than the others, are invalidated and considered as missing data. In the following, we  note this mask of valid slopes: 
\begin{equation}
        \label{eq:weight_valid}
        \weightvalid\Paren{\Vx} = 
        \begin{cases}
                1 \text{ if the slope at~$\Vx$ is valid,}
                \\
                0 \text{ otherwise.}
        \end{cases}
        \,
\end{equation}
As emphasised in black in \refsubfig{fig:2D_corr_sim_misreg}{a} and compared to the modelled slopes in \reffig{eq:weight_WFS}, this corresponds to an additional missing ring of one subaperture width on the outer ring and the corners of the inner ring of the pupil.

\begin{figure}[t!] % fig:2D_corr_sim_misreg
        \centering
        
        % Internal command of the figure for the automatic sizing
        % Path of the files
        \newcommand{\PathFig}{Fig_2D_corr_sim_misreg/}
        
        % Line ratio
        \newcommand{\LineRatio}{0.975}
        
        % Font of the text in the figure
        \newcommand{\fontTxt}[1]{\textbf{\tiny #1}}
        
        % Vertical space between lines
        \newcommand{\spaceLine}{-0.1cm}
        
        \newcommand{\subfigColor}{black}
        
        % Suffix KL
        \newcommand{\figOne}{\PathFig IM_shifted_KL_\numKL}
        \newcommand{\figTwo}{\PathFig IM_2D_corr_KL_\numKL}
        \newcommand{\figThree}{\PathFig 2D_corr}
        \newcommand{\figFour}{\PathFig 2D_corr_upsamp}
        
        % Getting the size of the boxes
        \sbox1{\includegraphics{\figOne}}
        \sbox2{\includegraphics{\figTwo}}
        \sbox3{\includegraphics{\figThree}}
        \sbox4{\includegraphics{\figFour}}
        
        % Defining column width command
        \newcommand{\ColumnWidth}[1]
                {\dimexpr \LineRatio \linewidth * \AspectRatio{#1} / (\AspectRatio{1} + \AspectRatio{2} + \AspectRatio{3} + \AspectRatio{4}) \relax
                }
        \newcommand{\ColumnGap}{\hspace {\dimexpr \linewidth /5 - \LineRatio\linewidth /5 }}

        % Figure table
        \begin{tabular}{
                @{\ColumnGap}
                M{\ColumnWidth{1}}
                @{\ColumnGap}
                M{\ColumnWidth{2}}
                @{\ColumnGap}
                M{\ColumnWidth{3}}
                @{\ColumnGap}
                M{\ColumnWidth{4}}
                @{\ColumnGap}
                }
                
                % Title line
                \fontTxt{Measured IM}
                &
                \fontTxt{Modal correlation}
                &
                \fontTxt{$\V{\alpha}$-map}
                &
                \fontTxt{Over-sampling}
                \\                
                
                \subfigimg[width=\linewidth,pos=ul,font=\fontfig{white}]{$\,$(a)}{0.0}{\figOne} &
                \subfigimg[width=\linewidth,pos=ul,font=\fontfig{\subfigColor}]{$\,$(b)}{0.0}{\figTwo} &
                \subfigimg[width=\linewidth,pos=ul,font=\fontfig{\subfigColor}]{$\,$(c)}{0.0}{\figThree} &
                \subfigimg[width=\linewidth,pos=ul,font=\fontfig{\subfigColor}]{$\,$(d)}{0.0}{\figFour}
                \\
        \end{tabular}        
               
        \caption{\label{fig:2D_corr_sim_misreg} Simulated example of the spatial 2D correlation method. \refpan{a}: simulation of a measured and shifted modal IM ($x$-slopes) for $\mKL=\numKL$. The grey and dark pixels correspond to the non-valid slopes~$\weightvalid\Paren{\Vx}=0$ in the measured IM.  \refpan{b}: cross-correlation of the shifted mode of \refpan{a} with the reference modal IM for $\mKL=\numKL$ (see \reffig{fig:2D_corr_KL}). \refpan{c}: map of the fitted coefficients $\V{\alpha}$. \refpan{d}: over-sampling of \refpan{c} with an interpolation via sinc functions.}
\end{figure}

In the following, we note $\IM{m}\Paren{\Vx}$ (resp. $\IMtilde{m}\Paren{\Vx}$) the reference modal IM of the system without any mis-registration (resp. the measured modal IM). Here, $\Vx$ is the 2D position in terms of subaperture of the SH-WFS and $m$ is the index of the considered mode. {The reference IM can be an IM measured during calibration on a well-aligned system, relaxing the need of a synthetic IM. For GPAO, we chose to use a synthetic IM which is computed once for all.}

For a given mode $m$, the cross-correlation  of the 2D IMs is given by:
\begin{equation}
    \label{eq:cross_corr}
    \Brack{\IMtilde{m}\crosscorr\IM{m}}\Paren{\Vdelta}
    {} \triangleq {} 
    \sum_{\Vx}\IMtilde{m}\Paren{\Vx}\IM{m}\Paren{\Vx-\Vdelta}
    {} = {}  \alpha\Paren{\Vdelta}
    \,.
\end{equation}
In terms of~$\Vdelta$, \refeq{eq:cross_corr} would give maps similar to the third line of \reffig{fig:2D_corr_KL} or \refsubfig{fig:2D_corr_sim_misreg}{b}. These maps can quickly be obtained by performing the cross-correlation in the Fourier space. Then, finding the value of~$\Vdelta$ maximising this correlation~$\alpha\Paren{\Vdelta}$ could give a hint on the lateral error. Nonetheless, as discussed above, the sensitivity of this solution will depend on the chosen mode~$m$ and this modal estimator does not provide a natural way to combine the different modes into a single general estimator. In addition, such a cross-correlation assumes that all the summed 2D positions~$\Vx$ are relevant. It does not account for the valid subapertures in the model nor in the measures via~\refeqs{eq:weight_WFS}{eq:weight_valid}% \HBO{je n'ai compris la phrase précédente qu'après avoir lu l'équation (6)}\ABE{J'ai modifié, est-ce mieux ?}
. As a consequence, this will strongly bias the estimated error in case of large mis-registrations due to pupil truncation effects.

In our method, we rather define~$\alpha\Paren{\Vdelta}$ in terms of how similar the two IMs are in terms of mean squares, jointly accounting for all the modes,
\begin{equation}
    \label{eq:2D_mod_min}
    \alpha\Paren{\Vdelta} = \argmin{\beta}{\sum_{\Vx,m}w\Paren{\Vx,\Vdelta}\Paren{\IMtilde{m}\Paren{\Vx} - \beta\IM{m}\Paren{\Vx-\Vdelta}}^2}
    \,,
\end{equation}
with:
\begin{equation}
    w\Paren{\Vx,\Vdelta} = \weightvalid\Paren{\Vx}\weightWFS\Paren{\Vx-\Vdelta}
    \,.
\end{equation}
The weight factor~$w\Paren{\Vx,\Vdelta}$ discards from the cost function the invalid pairs $\brack{\IMtilde{m}\Paren{\Vx},\IM{m}\Paren{\Vx-\Vdelta}}$ if one of the elements is estimated on an invalid position. \refeqfull{eq:2D_mod_min} has an analytical solution given by
\begin{align}
    \alpha\Paren{\Vdelta} {}={} & \frac{\sum_{\Vx,m}w\Paren{\Vx,\Vdelta}\IMtilde{m}\Paren{\Vx}\IM{m}\Paren{\Vx-\Vdelta}}{\sum_{\Vx,m}w\Paren{\Vx,\Vdelta}\Brack{\IM{m}\Paren{\Vx-\Vdelta}}^2}
    ,\\
    \label{eq:2D_corr_map}
    {}={} & \frac{\sum_{m}\Brack{\weightvalid\IMtilde{m}\crosscorr\weightWFS\IM{m}}\Paren{\Vdelta}}{\sum_{m}\Brack{\weightvalid\crosscorr\weightWFS\Brack{\IM{m}}^2}\Paren{\Vdelta}}
    \,.
\end{align}
Thus, the coefficient map~$\alpha\Paren{\Vdelta}$ is expressed by the ratio of the modal sum of 2D cross-correlations.

To achieve super-resolution without the need of a complex model fitting method, this low resolution map~$\alpha\Paren{\Vdelta}$ is further up-sampled by zero-padding its Fourier transform by a factor~$\mUS=8$. This factor ensures a resolution of \percent{12.5} of a subaperture, finer than the target error of \percent{20} mentioned in \refsec{sec:intro}. This step does not add any information, but it is equivalent to an interpolation with a sinc function, oversampling the maximum region with a continuous function. Other methods involving polynomial fit or an iterative weighted center of gravity could be used but slower and prone to converge to local maxima. On the contrary, using a zero-padding operation is numerical very efficient, and handle all maxima at once. In the end, the position of the global maximum of the obtained map gives the estimated lateral mis-registration~$\tilde{\Vdelta}$ at the resolution given by the up-sampling parameter~$\mUS$ in a single pass:
\begin{equation}
    \label{eq:2D_corr_estim}
    \tilde{\Vdelta} = \argmax{\Vdelta}\alpha\Paren{\Vdelta}
    \,.
\end{equation}
Looking closer at the definition of~$\alpha$ in~\refeq{eq:2D_mod_min}, it appears that~$\alpha\Paren{\tilde{\Vdelta}}$ gives the optimal coefficient to maximise the similarity between $\IM{m}$ and $\IMtilde{m}$ in the sense of the mean squares. In other words, $\alpha\Paren{\tilde{\Vdelta}}$ gives the correction to apply on the amplitude of the guessed PSIM to obtain the real global amplitude of the measured IM.

In the following, we pragmatically chose to use the set of the 50 first KL modes. As discussed in \refsec{sec:conclu_2D_corr_estim} and \refapp{app:KL_modes}, it was beyond the scope of this paper to further optimise the considered modes. The chosen option produces a modal base that can be shared with all the GPAO modes (\X{9}, \X{30} and \X{40} SH-WFS). In the following, we work within the framework of the \X{40} SH-WFS GPAO mode.

\subsubsection{Advantages of the spatial correlation estimator}

Here, we list the main advantages of the proposed method. First, by accounting for the valid subapertures with an appropriate map~\refeq{eq:weight_WFS}, this method is robust to large errors and strongly misaligned systems. It has a large capture range because of the exhaustive exploration of the parameter space permitted by \refeq{eq:2D_corr_estim}. The estimator is absolute. It provides unbiased and super-resolved measures, up to the resolution of the up-sampling parameter~$\mUS$.

Afterwards, the application of the method is fast: by focussing only on a limited numbers of modes, the acquisition of the modal IM is quick to obtain. And it is fast to compute: the convolutions in \refeq{eq:2D_corr_map}, with proper zero-padding, can be computed using fast 2D discrete Fourier transforms and the fit of the lateral error is obtained in a single pass via \refeq{eq:2D_corr_estim}. Thus, there is no need to recompute several PSIM to iterate the parameter fit.

Furthermore, the need of a PSIM model is reduced to the minimal need: a single computation to get the reference IM from which the command matrix of the AO loop is derived. For complex systems where a PSIM model is not available, this need could be removed by using a measured IM obtained during calibration on an aligned system.

Because it is based on lots of redundant measures to retrieve only the two parameters of the lateral error, the proposed estimator has a high S/N. This makes it particularly suitable for open-loop calibration where the turbulence strongly corrupts the slopes of the measured modal IM, as far as the measurement strategy freezes, in average, the turbulence~\citep{Kolb:16_SPIE_Review_AO_calibration}. In addition, \citet{Heritier:21_SPRINT} has shown that these kinds of approaches are not biaised by the turbulence wind.

Finally, an interesting by-product of the method is the global amplitude of the IM, which is given by the amplitude of the correlation peak. Although this does not provide the individual amplitude on each actuator of the DM, this is an interesting parameter for the monitoring and re-calibration of the system and its AO loop. Even if this beyond the scope of this paper which focuses on SH-WFS, we also notice here that estimating the amplitude of the IM could be trickier with a pyramid WFS as the optical gains will vary with the considered modes. This would change their relative weighting in \refeq{eq:2D_corr_map}, but should not modify the definition of the best position.

\subsection{Simulations: large capture range}
\label{sec:simu_2D_corr}

The modal correlation estimator was tested with the AOF tool embedded in SPARTA in charge of generating PSIMs. As described by \citet{Kolb:12_SPIE_AOF}, it uses a geometrical model of the SH-WFS. The slopes are obtained by computing the wavefront phase difference on the opposite edges of each subaperture. For the DM, we used a pseudo-synthetic model. For each actuator~$a$ of the ALPAO DM, its influence function~$\psi_a$ is described with a symmetrically radial profile:
\begin{equation}
    \label{eq:IF}
    \psi_a\Paren{\x} = A_a.e^{-\alpha_a \Paren{\Abs{\x}/\DMpitch}^{\beta_a}}
    \,,
\end{equation}
where $\DMpitch$ is the actuator pitch, where $A_a$ is the amplitude of the actuator, and where $\alpha_a$ and $\beta_a=$ are the parameters of the super-Gaussian profile. The values of these parameters are fitted on the real influence functions measured and kindly provided by ALPAO. The actuator positions and influence functions are stretched according the GPAO design\footnote{In each instance of GPAO, the DM, placed in the Coudé train of its UT, is tilted with respect to the incident beam by an angle of $\sim\SI{13.3}{\degree}$, leading to a stretch of $\sim\percent{2.7}$ \citep{LeBouquin:23_GPAO_design} and $\DMpitch\neq\SApitch$.} and consequently do not lie on a Fried geometry~\citep{Fried:77}. The KL modes are defined on the actuator command space, following the covariance method described by \citet{BertrouCantou:22_KLmodes}. 

To test the method, a zonal PSIM was generated with extreme lateral registration error of~$\Vdelta=\Paren{\delta_{x}, \delta_{y}}=\Paren{13.35, 8.65}\SApitch$ and an amplitude of \SI{4}{\micro\meter} for the DM actuators in the visible NGS mode of GPAO (\X{40} SH-WFS). Noisy measurements were simulated by adding a centered Gaussian noise on the slopes with a standard deviation of $\sigma^\Tag{slope}=0.25$ pixel, corresponding to a noise of $200$~mas in the GPAO design. Results are shown in \reffig{fig:2D_corr_sim_misreg}, using KL modes from $\mKL=4$ to $m^{\Tag{KL, max}}=50$.

\refsubfigfull{fig:2D_corr_sim_misreg}{a} shows the $\mKL=\numKL^\Tag{th}$ mode of the modal IM \mbox{obtained} after projecting the zonal IM on the KL modes. Compared with its centered representation in \reffig{fig:2D_corr_KL}, the introduced shift is visible with a slope pattern clearly off-centered from the WFS pupil. In addition, it appears that the noise on the zonal IM propagates towards the modal IM. For information, the correlation of this shifted noisy pattern with its reference is given in \refsubfig{fig:2D_corr_sim_misreg}{b}. The results is smooth, but finding the position of the maximal correlation is ambiguous with several potential locations (red regions).

This ambiguity is lifted when looking at the map of the fitted coefficients $\V{\alpha}$ in \refsubfig{fig:2D_corr_sim_misreg}{c}. Only one clear location emerges for the maximal correlation. The noise on this map is negligible and its oversampled version of \refsubfig{fig:2D_corr_sim_misreg}{d} leads to an estimated shift of~$\tilde{\Vdelta}=\Paren{13.375, 8.625}\SApitch$ and an amplitude of \SI{3.99}{\micro\meter}. The results is in the resolution of~$1/\mUS=0.125$ of a subaperture. The amplitude is also correctly estimated, within a percent.

% xShift: 1337.500000 (vs 1335.000000)
% yShift: 862.500000 (vs 865.000000)
% IM_amp: 3.992371 (vs 4.000000)

For information, the impact of the number of modes on the estimator performances is briefly discussed in \refapp{app:KL_modes}. As studied in \refapp{app:cross_talk_2D_corr}, the cross-talk with other mis-registration parameters (rotation and magnification) is negligible for realistic cases.

\subsection{Experimental results: System alignment at preset}
\label{sec:exp_2D_corr}

\begin{figure}[t!]
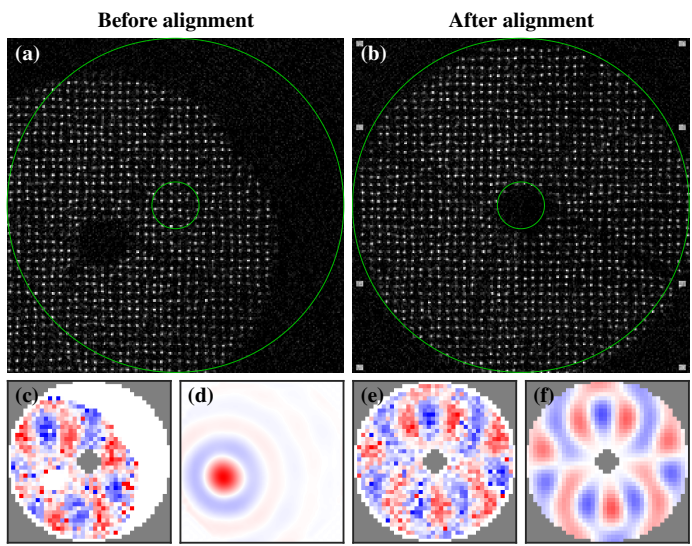
 % fig:2D_corr_bench
    \centering
        
    % Internal command of the figure for the automatic sizing
    % Path of the files
    \newcommand{\PathFig}{Fig_2D_corr_bench/}
    
    % Vertical space between lines
    \newcommand{\spaceLine}{-0.1cm}
    % Font of the text in the figure
    \newcommand{\fontTxt}[1]{\textbf{\tiny #1}}
    
    % Line ratio
    \newcommand{\LineRatio}{0.99}
    % Color of the panel letter
    \newcommand{\subfigColor}{white}
    % Defining column width command
    \newcommand{\ColumnWidth}
        {\dimexpr \LineRatio \linewidth /2 \relax
        }
    \newcommand{\ColumnGap}{\hspace {\dimexpr \linewidth /1 - \LineRatio\linewidth /1 }}
    % Figure table
    \begin{tabular}{
        @{}
        M{\ColumnWidth}
        @{\ColumnGap}
        M{\ColumnWidth}
        @{}
        }
        
        % Title line
        \fontTxt{Before alignment}
        &
        \fontTxt{After alignment}
        \\                
        \subfigimg[width=\linewidth,pos=ul,font=\fontfig{\subfigColor}]{$\,$(a)}{0.0}{\PathFig Pixel_START} &
        \subfigimg[width=\linewidth,pos=ul,font=\fontfig{\subfigColor}]{$\,$(b)}{0.0}{\PathFig Pixel_END}
    \end{tabular}
    
    \vspace{\spaceLine}
    
    % Line ratio
    \renewcommand{\LineRatio}{0.97}
    % Color of the panel letter
    \renewcommand{\subfigColor}{black}
    % Defining column width command
    \renewcommand{\ColumnWidth}
            {\dimexpr \LineRatio \linewidth /4 \relax
            }
    \renewcommand{\ColumnGap}{\hspace {\dimexpr \linewidth /3 - \LineRatio\linewidth /3 }}
    % Figure table
    \begin{tabular}{
        @{}
        M{\ColumnWidth}
        @{\ColumnGap}
        M{\ColumnWidth}
        @{\ColumnGap}
        M{\ColumnWidth}
        @{\ColumnGap}
        M{\ColumnWidth}
        @{}
        }
        
        % Title line
        \subfigimg[width=\linewidth,pos=ul,font=\fontfig{\subfigColor}]{$\,$(c)}{0.0}{\PathFig IM_modal_KL_\numKLbench_START} &
        \subfigimg[width=\linewidth,pos=ul,font=\fontfig{\subfigColor}]{$\,$(d)}{0.0}{\PathFig 2D_corr_START} &
        \subfigimg[width=\linewidth,pos=ul,font=\fontfig{\subfigColor}]{$\,$(e)}{0.0}{\PathFig IM_modal_KL_\numKLbench_END} &
        \subfigimg[width=\linewidth,pos=ul,font=\fontfig{\subfigColor}]{$\,$(f)}{0.0}{\PathFig IM_synth_KL_\numKLbench_END}
    \end{tabular}
    
    \caption{\label{fig:2D_corr_bench} Example in the GPAO bench of the 2D modal estimator. \refpans{a,b}: SH-WFS pixels before and after the alignment. Green circles: WFS pupil edges. \refpan{c,e,f}: $x$-slopes of the modal IM for $\mKL=\numKLbench$ before (\refpan{c}) and after (\refpan{e}) alignment and reference (\refpan{f}). \refpan{d}: over-sampled map of the coefficient $\V{\alpha}$ before alignment.}
\end{figure}

\begin{figure}[t!]
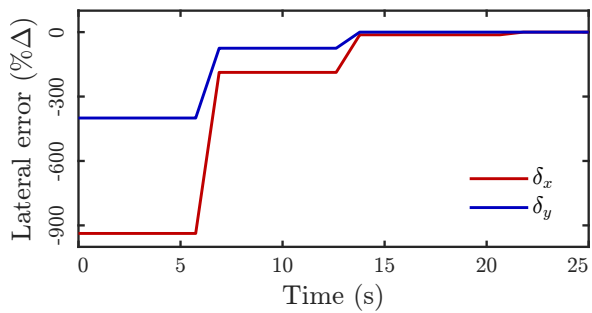
 % fig:2D_corr_bench_conv
    \centering
    
   % Internal command of the figure for the automatic sizing
    % Path of the files
    \newcommand{\PathFig}{Fig_2D_corr_bench/}
    
    \subfigimg[width=0.85\linewidth,pos=ul,font=\fontfig{black}]{}{0.0}{\PathFig Conv_curv}

	\caption{\label{fig:2D_corr_bench_conv} {Convergence of the lateral error corrective loop with the 2D modal estimator at GPAO preset.}}
\end{figure}

The 2D modal estimator was tested in the GPAO development bench. To show that the method does not rely on a Fried geometry, an angle of \SI{35}{\degree} was applied between the DM pupil and the WFS pupil. The wind was emulated by spinning the phase plate at its maximal speed. This produces a wind of $\vWind\simeq\SI{8.4}{\meter\per\second}$ with a Fried parameter of~$\rFried \simeq \SI{14}{\centi\meter}$. The WFS stage was translated by several subapertures, as illustrated in \reffig{fig:2D_corr_bench}{a}. A few iterations of the lateral error corrective loop were performed with a gain of 0.8 as shown in \reffig{fig:2D_corr_bench_conv}. To freeze the turbulence, the modal IMs were measured with the fast `push-pull' method \citep{Kasper:04_Hadamard, Oberti:06_SPIE_PSIM_vs_measured_IM, Kolb:16_SPIE_Review_AO_calibration}, playing sequentially the different modes. The AO loop was open.

Qualitatively speaking, the shift of the pupil in \refsubfig{fig:2D_corr_bench}{a} is clearly visible in the modal IM of \refsubfig{fig:2D_corr_bench}{c} despite the high level of noise. The under-illumination threshold of SPARTA automatically sets the slopes of poorly illuminated subapertures to zero. The over-sampled map of \refsubfig{fig:2D_corr_bench}{d} has a clear optimal area.

The corrective loop converges in only three iterations, as seen in \reffig{fig:2D_corr_bench_conv}. The period of the iteration is dominated by the overhead times of SPARTA when measuring the IM and the time to move the actuator of the translating stage. The time to actually poke the DM is \SI{10}{\milli\second} per mode, so half a second, and the time to correlated the measured IM with the model is negligible. After the alignment, the noisy pattern of the modal IM of \refsubfig{fig:2D_corr_bench}{e}, convingly matches the reference pattern of \refsubfig{fig:2D_corr_bench}{f}. Furthermore, it can be seen from \refsubfig{fig:2D_corr_bench}{b} that the photometric pupil of the system is off-centred\footnote{As a side remark, the eight white squares in \reffig{fig:2D_corr_bench}{b} come from the activation during the instrument calibration of the dark follower of the eight octants of the camera.}. Some subapertures do not get flux despite being in the SH-WFS pupil (top-left) while some others outside the pupil are illuminated (bottom-right). This supports the fact that except for a rough alignment of the system, the photometric pupil cannot be used to align the DM actuator geometry and the SH-WFS. 

To quantitatively assess the alignment efficiency, interaction matrices were measured prior and after the system auto-alignment. For these measurements, the phase plate was stopped to prevent any bias. SPARTA embeds an AOF tool to estimate the mis-registration parameters of the system by PSIM iterative model fitting, as described by \citet{Kolb:12_SPIE_AOF}. The results are presented in \reftab{tab:2D_corr_bench}.

\begin{table}[t!] % tab:2D_corr_bench
    \caption{\label{tab:2D_corr_bench} System auto-alignment with the 2D modal estimator}
    \centering
    \begin{tabular}{ccccc}
    \hline
    \hline
     & Amplitude & $\delta_x~/~\delta_y$ & $\theta$ & $\delta_{\rho_x}~/~\delta_{\rho_y}$
    \\
     & (\SI{}{\micro\meter}) & ($\%\SApitch$) & (\SI{}{\degree}) & ($\%$)
    \\
    \hline
    $t=\SI{0}{\second}$ & 3.6 & -908.5~/~-377.4 & 35.5 & 0.8~/~1.4
    \\
    $t=\SI{25}{\second}$ & 3.9 & 0.9~/~-3.6 & 35.2 & 0.6~/~1.3
    \\
    \hline
    \end{tabular}
    \tablefoot{PSIM parameters fitted by SPARTA on the IM measured before ($t=\SI{0}{\second}$) and after ($t=\SI{25}{\second}$) the system auto-alignment with the 2D modal estimator. $\theta$: clocking. $\delta_{\rho_x}~/~\delta_{\rho_y}$: stretches along the $x$ and $y$-axes.}
\end{table}

% SL START: -937.5000 -400.0000
% SL END: amplitude = 3.3920

% ----- START -----
% Variable / 2D_corr / SPARTA
% Amplitude: 3.256041e+00 / 3.622906e+00
% xShift: -950 / -9.085130e+02
% yShift: -400 / -3.774327e+02
% Rotation: 215 / 2.154922e+02
% xStretch: 0 / 7.565936e-01
% yStretch: 0 / 1.379383e+00
% ----- START -----

% ----- END -----
% Variable / 2D_corr / SPARTA
% Amplitude: 3.522174e+00 / 3.926853e+00
% xShift: 0 / 8.864518e-01
% yShift: -1.250000e+01 / -3.587397e+00
% Rotation: 215 / 2.152492e+02
% xStretch: 0 / 5.981908e-01
% yStretch: 0 / 1.283499e+00
% ----- END -----

The initial lateral error is~$\Vdelta=\Paren{-908.5,-377.4}\%\SApitch$. From \reffig{fig:2D_corr_bench_conv}, with the phase plate spinning, the error initially found by the 2D modal estimator is~$\tilde{\Vdelta}=\percent{\Paren{-937.5,-400.0}}\SApitch$. Remembering that~$\mUS=8$, this estimation lies within three resolution elements of $\percent{12.5}\SApitch$ of the up-sampled map $\V{\alpha}$. This is thus better than half a subaperture and despite the fact that the system presents a magnification of~$\sim\percent{1}$, not considered in the estimator. After convergence ($\tilde{\Vdelta}=\V{0}$), \reftab{tab:2D_corr_bench} shows that the actual lateral misalignment is well below the up-sampled map $\V{\alpha}$ resolution, with a residual of a few percent of a subaperture. 

At convergence, the amplitude fitted by the 2D modal estimator is $\SI{3.4}{\micro\meter}$. If this is the correct order of magnitude, it is $\sim\percent{13}$ below the amplitude given by the SPARTA fit given in \reftab{tab:2D_corr_bench}. Several factors can explain this discrepancy. First, the IM is measured by playing a Hadamard set of zonal commands~\citep{Kasper:04_Hadamard, Oberti:04_SPIE_IM_misreg}, producing potentially strong local slopes on the DM and thus not the same linearity point of the system than the one of the modal IM measurement. In addition, the~$\Abs{\delta_\rho}\simeq\percent{1.5}$ stretch in the system measured by SPARTA is not included in the reference IM of the estimator. This can further bias the model fitting of \refeq{eq:2D_corr_estim}. Better understanding this amplitude was not considered critical in this study. This is indeed an interesting by-product of the method, but we mainly wanted to focus on the system auto-alignment.

Finally and importantly, we remark here that the precision of the alignment obtained with the 2D modal estimator is sufficient to close the AO loop with an IM and a control matrix synthesised for a perfectly aligned system ($\Vdelta=\V{0}$) and controlling 500~modes, which corresponds to the baseline of GPAO for standard operation. This shows that the 2D modal estimator is a robust tool to put the system in state where the closed loop estimator presented hereafter can take over the lateral misalignment monitoring and correction.

\section{Non-perturbative closed loop estimator}
\label{sec:main_CL_estim}

\subsection{Proposed method}
\label{sec:method_CL_estim}

\subsubsection{Block diagram of an AO loop}

A standard AO loop works as follows. First, the WFS~$S$ converts the 2D wavefront~$\Vw$ into measurements~$\Vm$ corrupted by noise,~$\Vn$:
\begin{equation}
        \Vm = S\Paren{\Vw} + \Vn
        \,.
\end{equation}
Next, these measurements are processed by the controller~$C$ to compute a new command to send to the DM:
\begin{equation}
        \Vc = C\Paren{\Vm}
        \,.
\end{equation}
Then, this command is applied and held by the DM until the next command arrives. For example, in the case of GRAVITY+, $\Vm$ are slopes delivered by the SH-WFS and the controller is a leaky integrator\footnote{We notice here the possibility that~$\gint$ and~$\gleak$ could be functions applied on~$\Vc^{i}$ and~$\Vc^{\delta}$ and not pure scalars, especially if modal filtering is applied for control and leakage \citep{Gendron:94_AA_Modal_control}.} of leak gain~$\gleak$ and integral gain~$\gint$:
\begin{equation}
        \Vc^{i+1} = \Paren{1-\gleak}\Vc^{i} + \gint\Vc^{\delta}
        \,,
\end{equation}
and where the command correction:
\begin{equation}
        \Vc^{\delta} = \CM \Vm
        \,,
\end{equation}
is obtained with a command matrix~$\CM$ computed by inverting the PSIM of the system filtered on a given~$\nmodes$ number of KL modes.

\begin{figure}[t!] % fig:CL_corr_spatial_coupling
        \centering
        
        % Internal command of the figure for the automatic sizing
        % Path of the files
        \newcommand{\PathFig}{Fig_CL_corr_spatial_coupling/}
        
        \includegraphics[width=\linewidth]{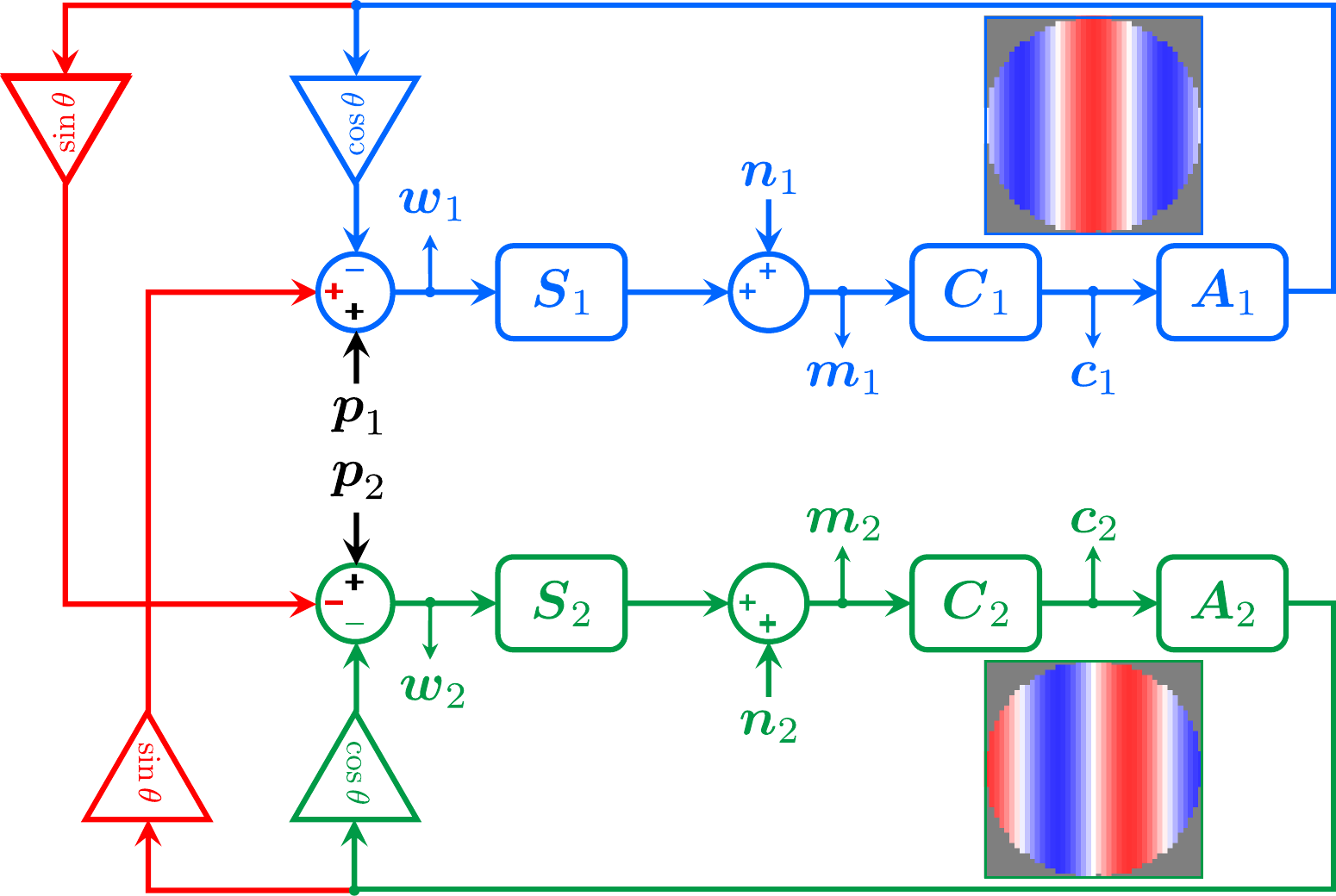}

        \caption{\label{fig:CL_corr_spatial_coupling} Block diagram of the temporal coupling (in red) of the symmetric (cosine, in blue) and anti-symmetric (sine, in green) parts of a given spatial frequency~$\Vk=\Paren{2/3D, 0}$ through the AO loop.}
\end{figure}

In the following, we work under different assumptions. Firstly, when an AO system is in closed loop, it can be linearised around its operating point, and all the previously described steps become linear operations. It is then equivalent to describe the AO loop in the Fourier space of the commands~$\Vm$ and of the wavefront~$\Vw$ that is to say in terms of 2D spatial frequencies~$\Vk=\Paren{k_x, k_y}$~(\SI{}{\per\meter}) propagating through the system. Secondly, in such a system, without any misalignment (no lateral shift, no clocking and no stretch between the sensor and the actuators), there is no cross-coupling between different 2D spatial frequencies~$\Vk\neq\Vk^\prime$. For a spatial frequency~$\Vk$, the above steps (i), (ii), and (iii) can then be represented by the block diagram of \reffig{fig:CL_corr_spatial_coupling} where each block is a linear operation. The symmetric (equivalently the real part of the Fourier coefficient) and the anti-symmetric part (equivalently the imaginary part of the Fourier coefficient) of this spatial frequency are presented with the subscripts one (blue) and two~(green).

The effect of the different blocks of \reffig{fig:CL_corr_spatial_coupling} can be described by their transfer function defined in the temporal frequency space~$f$ (\SI{}{\hertz})~\citep{Astrom:21_Feedback_system}. It is a reasonable assumption to consider that the transfer functions are identical for the two spatial modes:
\begin{equation}
    \label{eq:eq_12}
    \begin{cases}
    S_{1}=S_{2}=S
    \,,
    \\
    C_{1}=C_{2}=C
    \,,
    \\
    A_{1}=A_{2}=A
    \,.
    \end{cases}
\end{equation}

As detailed by~\citet{Madec:99_control}, the WFS integrates the signal during its exposure time~$\tWFS$ giving the transfer function:
\begin{equation}
    % \fdep{S} = \frac{1-e^{-2i\pi\tWFS f}}{2i\pi\tWFS f}
    S\Paren{f} = \frac{1-e^{-2i\pi\tWFS f}}{2i\pi\tWFS f}
    \,.
\end{equation}
Then, $C$ is a leaky controller of transfer function:
\begin{equation}
    % \fdep{C} = \frac{\gint e^{-2i\pi\tlat f}}{1-\Paren{1-\gleak} e^{-2i\pi\tRTC f}}
    C\Paren{f} = \frac{\gint e^{-2i\pi\tlat f}}{1-\Paren{1-\gleak} e^{-2i\pi\tRTC f}}
    \,,
\end{equation}
where~$\tlat$ is the latency of the system (communication and computation times) and~$\tRTC$ is the period of the RTC cycle\footnote{Following a previous note, $\gint\Paren{\Vk}$ and $\gleak\Paren{\Vk}$ can depend on the spatial frequency~$\Vk$ if modal control is implemented.}. Finally, the DM acts as a zero-controller holder, maintaining the command during~$\tDM$ with a transfer function similar to the WFS:
\begin{equation}
    % \fdep{A} = \frac{1-e^{-2i\pi\tDM f}}{2i\pi\tDM f}
    A\Paren{f} = \frac{1-e^{-2i\pi\tDM f}}{2i\pi\tDM f}
    \,.
\end{equation}
In general, all the characteristic times are equal:
\begin{equation}
    \label{eq:tau_characteristic}
    \tlat \simeq \tDM \simeq \tWFS \simeq \tRTC
    \,.
\end{equation}

\subsubsection{Temporal coupling of a 2D spatial frequency}
\label{sec:temp_coupling}

In the following, $\Vx=\Paren{x,y}$ denotes for the 2D spatial position. Using the notations of \reffig{fig:CL_corr_spatial_coupling}, a lateral 2D shift~$\Vdelta=\Paren{\delta_x, \delta_y}$ between the DM and the WFS implies a spatial phase shift between the spatial frequency~$\Vk$ of the wavefront~$\Vw$ seen by the WFS and its correction~$\Vc$ by the DM as follows:
\begin{equation}
    \begin{cases}
        \Paren{1} 
        & \Vc_{1} \propto \cos\Paren{2\pi\Vk\scaprod\Vx}
        \Rightarrow
        \Vw {}\propto{} -\cos\Paren{2\pi\Vk\scaprod\Paren{\Vx-\Vdelta}}
        \,,
        \\
        \Paren{2}
        & \Vc_{2} {}\propto{} \sin\Paren{2\pi\Vk\scaprod\Vx}
        \Rightarrow
        \Vw {}\propto{} -\sin\Paren{2\pi\Vk\scaprod\Paren{\Vx-\Vdelta}}
        \,.
    \end{cases}
\end{equation}
Which, after expansion, leads to:
\begin{equation}
    \begin{cases}
        \Paren{1} 
        &
        \Vw {}\propto{}
        -\cos\Paren{\theta}\cos\Paren{2\pi\Vk\scaprod\Vx}
        -\sin\Paren{\theta}\sin\Paren{2\pi\Vk\scaprod\Vx}
        \,,
        \\
        &
        \Vw {}\propto{}
        -\cos\Paren{\theta}\Vw_{1}
        -\sin\Paren{\theta}\Vw_{2}
        \,,
        \\
        \Paren{2}
        &
        \Vw {}\propto{}
        +\sin\Paren{\theta}\cos\Paren{2\pi\Vk\scaprod\Vx}
        -\cos\Paren{\theta}\sin\Paren{2\pi\Vk\scaprod\Vx}
        \,,
        \\
        &
        \Vw {}\propto{}
        +\sin\Paren{\theta}\Vw_{1}
        -\cos\Paren{\theta}\Vw_{2}
        \,,
    \end{cases}
\end{equation}
with the coupling coefficient of:
\begin{equation}
    \label{eq:theta}
    \theta = 2\pi\Vk\scaprod\Vdelta
    \,.
\end{equation}
As emphasised by the red arrows in \reffig{fig:CL_corr_spatial_coupling}, there is now consequently a cross-coupling between the symmetric (1) and anti-symmetric (2) parts of the spatial frequency~$\Vk$.

The commands~$\Vc_{1}$ and~$\Vc_{2}$ are thus no longer independent. Discussing on the noise propagation through an AO loop, \citet[Sect. 4.1.3]{Heritier:19_PHD} already mentioned the fact that measurement noise can produce signals and intuited that the larger the mis-registration, the higher the S/N on the measures would be. The present method is based on the fact that such a coupling between~$\Vc_{1}$ and~$\Vc_{2}$ leaves a trace in their temporal correlation. In the (spatial and temporal) Fourier space, the variance of the command~$c_{i}\Paren{\Vk,f}$ of a given spatial frequency~$\Vk$ at a temporal frequency~$f$ is defined by:
\begin{equation}
    \label{eq:var}
    % \Vvar{\cf{i}}\Paren{\Vk,f} \triangleq \Avg{\cf{i}\Paren{\Vk,f}\cfconj{i}\Paren{\Vk,f}}
    \Vvar{\Vc_{i}}\Paren{\Vk,f} \triangleq \Avg{c_{i}\Paren{\Vk,f}\conj{c}_{i}\Paren{\Vk,f}}
    \,,
\end{equation}
and the correlation between~$\Vc_{1}$ and~$\Vc_{2}$ for this given spatial frequency~$\Vk$ at the temporal frequency~$f$ is defined by:
\begin{equation}
    \label{eq:corr}
    \Vcorr{\Vc_{1},\Vc_{2}}\Paren{\Vk,f} \triangleq \frac{\Avg{c_{1}\Paren{\Vk,f}\conj{c}_{2}\Paren{\Vk,f}}}{\sqrt{\Vvar{c_{1}}\Paren{\Vk,f}\Vvar{c_{2}}\Paren{\Vk,f}}}
    \,.
\end{equation}

\begin{figure}[t!] % fig:CL_corr_curves
    \centering
    
    % Internal command of the figure for the automatic sizing
    % Path of the files
    \newcommand{\PathFig}{Fig_CL_corr_curves/}
    
    \includegraphics[width=0.85\linewidth]{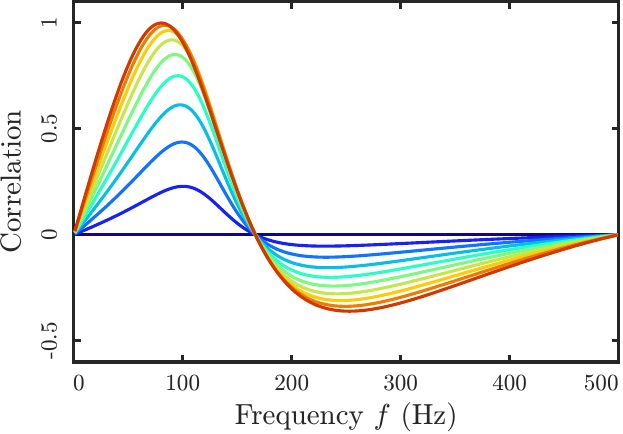}
    
    \caption{\label{fig:CL_corr_curves} Imaginary part of the correlation curves~$\Vcorr{\Vc_{1},\Vc_{2}}\Paren{\theta,f}$ in a noise limited regime for various values of $\theta$ from \SI{0}{\degree} (blue) to \SI{45}{\degree} (red), every~\SI{5}{\degree}. AO parameters: $\tWFS=\SI{1}{\milli\second}$, $\gint=0.5$, $\gleak=0$.}
\end{figure}

Under the assumption that the different noise terms~$\Vn_{i}$ and~$\Vp_{i}$ are independent, we show in \refapp{app:transfer_functions} that the coupling of $\Vc_{1}$ and $\Vc_{2}$ through the system produces a signature in their correlation that only depends on the coefficient~$\theta$ and the frequency~$f$. This signature is expressed by:
\begin{equation}
    \label{eq:corr_0}
    % \Vcorr{\cf{1},\cf{2}}\Paren{\Vk,\theta}  / \theta \underset{\theta\rightarrow0}{\sim} 2i\imag{\frac{\mufconj}{1+\mufconj}}
    \Vcorr{\Vc_{1},\Vc_{2}}\Paren{\theta\Paren{\Vk},f}  / \theta\Paren{\Vk} \underset{\theta\rightarrow0}{\sim} 2i\imag{\frac{\muconj\Paren{f}}{1+\muconj\Paren{f}}}
    \triangleq i\Vcorr{0}\Paren{f}
    \,,
\end{equation}
with:
\begin{equation}
    % \muf \triangleq \fdep{A}\fdep{C}\fdep{S}
    % \mu\Paren{f} \triangleq A\Paren{f}C\Paren{f}S\Paren{f}
    \mu \triangleq A C S
    \,.
\end{equation}
The correlation of a given spatial frequency~$\Vk$ is thus a pure imaginary number. \reffigfull{fig:CL_corr_curves} shows the imaginary parts of the correlation~$\Vcorr{\Vc_{1},\Vc_{2}}\Paren{\theta,f}$ for different values of the coupling coefficient~$\theta$.

\subsubsection{Estimation of the registration error from telemetry}

First, to apply the results of \refsec{sec:temp_coupling}, the spatio-temporal cube of commands~$c\Paren{\Vx,t}$ recorded in the closed loop telemetry must be reshaped in terms of spatial and temporal frequencies and split on its symmetric and anti-symmetric compounds. With the notation of \reffig{fig:CL_corr_spatial_coupling}, this 3D cube can be written as:
\begin{equation}
    c\Paren{\Vx,t} = \sum_{\Vk} c_{1}\Paren{\Vk,t} \cos\Paren{2\pi\Vk\scaprod\Vx} + c_{2}\Paren{\Vk,t} \sin\Paren{2\pi\Vk\scaprod\Vx}
    \,.
\end{equation}
with:
\begin{equation}
    c_{1}\Paren{-\Vk,t} = c_{1}\Paren{\Vk,t} \text{ and } c_{2}\Paren{-\Vk,t} = - c_{2}\Paren{\Vk,t}
    \,,
\end{equation}
to ensure the parity of the symmetric and anti-symmetric parts~$\Vc_{1}$ and~$\Vc_{2}$. As a consequence, the discrete 3D Fourier transform of~$\Vc$, $\TF{\Vc}\Paren{\Vk,f}$, is equal to:
\begin{equation}
    \TF{\Vc}\Paren{\Vk,f} = c_{1}\Paren{\Vk,f} - i c_{2}\Paren{\Vk,f}
    % \triangleq
    % \cf{1}\Paren{\Vk} - i \cf{2}\Paren{\Vk}
    \,,
\end{equation}
and thus:
\begin{equation}
    \begin{cases}
        c_{1}\Paren{\Vk,f} = \frac{1}{2} \Brack{\TF{\Vc}\Paren{\Vk,f} + \TF{\Vc}\Paren{-\Vk,f}}
        \,,
        \\
        c_{2}\Paren{\Vk,f} = \frac{i}{2} \Brack{\TF{\Vc}\Paren{\Vk,f} - \TF{\Vc}\Paren{-\Vk,f}}
        \,.
    \end{cases}
\end{equation}

Then, the imaginary part of the empirical correlation of the closed loop telemetry is computed as follows:
\begin{equation}
    \label{eq:corr_empirical}
    \corrt{cl}\Paren{\Vk,f} = \imag{\frac{c_{1}\Paren{\Vk,f}\conj{c}_{2}\Paren{\Vk,f}}{\Abs{c_{1}\Paren{\Vk,f}}\Abs{c_{2}\Paren{\Vk,f}}}}
    \,.
\end{equation}
The control space~$\Vk\in\kctrl$ of the AO loop is delimited by a disk of radius~$k^\Tag{max}$, in terms of cycle per diameter, so that:
\begin{equation}
    \label{eq:kctrl}
    \pi \Brack{k^\Tag{max}}^2 = \frac{\nmodes}{\nact}\brack{\dact}^2
    \,,
\end{equation}
where $\dact$ is the number of actuators across the diameter of the telescope and~$\nact$ is the total number of actuators of the DM. \refeqfull{eq:kctrl} states that among the~$\brack{\dact}^2$ modes in the Cartesian Fourier space, only the ratio of the number of controlled modes~$\nmodes$ over the maximal number of degree of freedom are controlled. We thus get from~\refeqs{eq:theta}{eq:corr_0} that:
\begin{equation}
    \forall \Vk\in\kctrl, \corrt{cl}\Paren{\Vk,f} \simeq 2\pi\Vcorr{0}\Paren{f}\Vk\scaprod\Vdelta
    \,.
\end{equation}

Finally, the lateral error is given by:
\begin{equation}
    \label{eq:CL_estim}
    \tilde{\Vdelta} = \argmin{\Vdelta}{\sum_{\Vk\in\kctrl, f>0}\Paren{\corrt{cl}\Paren{\Vk,f} - 2\pi\Vcorr{0}\Paren{f}\Vk\scaprod\Vdelta}^2}
    \,.
\end{equation}
This problem is solved by expanding the dimensions along~$\Vk$ and $f$ in order to reshape the equation into a matrix-vector shape,
\begin{equation}
    \corrt{cl}\Paren{\Vk,f} \simeq H\Paren{\Vk,f} \times \Vdelta
    \,,
\end{equation}
and using the pseudo-inverse $\V{H}^\dag$ of the obtained matrix \citep{Moore:20_pinv,Penrose:55_pinv},
\begin{equation}
    \tilde{\Vdelta} = \Brack{H\Paren{\Vk,f}}^\dag \times \corrt{cl}\Paren{\Vk,f}
    \,.
\end{equation}

% - "This problem is solved by expanding the dimensions along k and f in order to reshape the equation into a matrix-vector shape and using the pseudo-inverse of the obtained matrix.”
% -> Je suivais jusque la, mais je ne comprends pas cette phrase… Ca veut dire que tu peux écrire l’equation au dessus sous form matricielle et la résoudre en inversant la matrice? Pas sur de comprendre comment… peut etre peux tu expliciter?

\subsubsection{Advantages of the closed loop estimator}
\label{sec:CL_estim_advantages}

As discussed below, the proposed method has numerous advantages. Firstly, it is purely based on the geometry of the chosen observable which can be the WFS measurements~$\Vm_{i}$ or the DM commands~$\Vc_{i}$. It only uses the history of this observable and its correlation. Knowing explicitly the link between WFS measurements and the DM commands is not needed. Thus, the method does not rely on any PSIM model from which mis-registration parameters must be retrieved and the method is consequently not prone to model errors. In addition, such PSIM models are generally complex (and potentially non-invertible) and extremely slow to compute, strongly slowing the iterative fit of the alignment errors.
Secondly, the model only depends on only a small number of parameters driving the AO loop. These parameters are generally well constrained by design or proper calibration.

Furthermore, the estimator, expressed in the Fourier domain, is sparse. This makes it extremely fast to compute, a critical aspect when considering the ever increasing number of actuators and system complexity of the future ELTs.

Even if it based on noise propagation, assumptions on the noise model remain minimal. The method only supposes that the noise is spatially uncorrelated, but for a given spatial frequency, it does not assume a specific noise model. If white noises~$\Vn_{i}$ would excite all the spatial and temporal frequencies and thus produce a strong signal in the command correlation, knowing the power spectrum density of the noise sources is not needed as it cancels out via \refeq{eq:corr}, as shown in \refapp{app:transfer_functions}.

Finally, a common issue with lateral error estimators based on the AO telemetry is their sensitivity to the wind in the case of a frozen flow turbulence \citep{Heritier:19_frozen_wind, Heritier:19_PHD}. As discussed in \refapp{app:turb_prop}, such a situation breaks the assumption that~$\Vp_{1}$ and~$\Vp_{2}$ are independent, used to get \refeq{eq:corr_0}. Nonetheless, as far as there is noise injected in the loop to produce a correlation signal in the commands, this effect is sparse in the temporal frequency space. Thus, compared to other methods based on AO loop telemetry \citep{Bechet:12_SPIE, Kolb:12_SPIE_AOF}, the wind should have a limited impact on the proposed estimator of \refeq{eq:CL_estim}.

\subsection{Simulations: Sensitivity and wind bias}
\label{sec:simu_CL_estim}

\subsubsection{End-to-end simulations}

We used an ideal model for the DM. The influence functions~$\psi_a=\psi$ were set to be identical for all actuators~$a$ by \mbox{replacing} the parameters in \refeq{eq:IF} by their median value fitted on the ALPAO DM: $A_a=\SI{9}{\micro\meter}$, $\alpha_a=0.87$ and $\beta_a=1.31$. For a null lateral error, the actuators were placed on a \X{41} grid ($\dact=41$) in a Fried geometry~\citep{Fried:77} leading to $\nact = 1353$ active actuators. Thus the DM pitch $\DMpitch$ is also equal to the size of a subaperture $\SApitch$. To introduce a minimal control on loop divergence, the commands are clipped between minus one and one.

A purely geometrical model was used to simulate the SH-WFS. The slopes were estimated by averaging the gradient obtained by finite difference in each subaperture of the SH-WFS at the wavelength~$\lambda^\Tag{wfs}=\SI{750}{\nano\meter}$. There is no diffractive propagation of the wavefront to produce spot and consequently no spot centroiding. Only the photon shot noise was considered and added directly to the geometrical slopes, using \refeq{eq:sig_ph} as detailed in \refapp{app:slope_noise}. 

\reftab{tab:list_variable} sums up the different notations used in \refsecs{sec:simu_CL_estim}{sec:exp_CL_estim}. The sensitivity of the estimator to some of these parameters, especially the wind speed~$\vWind$ and the noise level~$\nph$, is studied in the following, only focussing on the lateral error $\Vdelta$ introduced in the system. Cross-talk with other mis-registration is discussed in \refapp{app:cross_talk_CL_corr}.

\begin{table}[t!] % tab:list_variable
    \caption{\label{tab:list_variable} List of the main variables used in \refsecs{sec:simu_CL_estim}{sec:exp_CL_estim}.}
    \centering
    \begin{tabular}{cc}
    \hline
    \hline
    Variable & Description
    \\
    \hline
    $\Vdelta$ & \makecell{2D lateral error of Cartesian coordinates   \\ $\Paren{\delta_{x}, \delta_{y}}$  and polar coordinates       $\Paren{\Abs{\Vdelta}, \Arg{\Vdelta}}$}
    \\
    $\SApitch$ & \makecell{
        Size of a subaperture (40 accross \\ the pupil diameter)
        }
    \\
    $1/\tWFS=\SI{1}{\kilo\hertz}$ & \makecell{Frequency of the AO loop with \\ $\tlat \simeq \tDM \simeq \tWFS \simeq \tRTC$,  \refeq{eq:corr_t}}
    \\
    % $\tlat \simeq \tDM \simeq \tWFS \simeq \tRTC$ & \refeq{eq:tau_characteristic}
    % \\
    $\nmodes=500$ & \makecell{Number of controlled modes \\ (GPAO baseline)}
    \\
    $500$ & \makecell{Number of frames in the \\ telemetry batch (\SI{0.5}{\second})}
    \\
    $\vWind$ & Wind speed of the frozen flow
    \\
    $\thetaWind$ & Orientation of the frozen flow
    \\
    $\nph$ & The number of photons per subaperture
    \\
    $\gint=0.5$ & Integral gain of the AO loop
    \\
    $\gcl$ & \makecell{Sensitivity of the closed loop \\ estimator, \refeq{eq:gain_cl}}
    \\
    $\Thong = 0.5$ & \makecell{Gain of the lateral error corrective  \\ loop, \refeq{eq:corrective_loop}}
    \\
    $\Vcorr{0}$ & Theoretical correlation curve, \refeq{eq:corr_0}.
    \\
    $\corrt{t}$ & \makecell{Best fit curve of the temporal \\ correlation, \refeq{eq:corr_t}}
    \\
    $\corrt{2D}$ & \makecell{Best 2D fit map of the correlation \\ coefficients \refeq{eq:corr_2D}}
    \\
    \hline
    \end{tabular}
    \tablefoot{Excepted when stated otherwise, the values given in this table are kept constant through all the simulations and experiments.}
\end{table}

\subsubsection{Sensitivity of the closed loop estimator}

\label{sec:CL_estim_sim_gain}

\begin{figure}[t!]
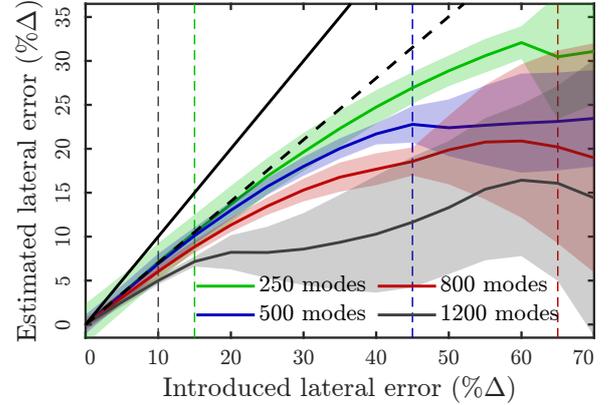
 % fig:CL_corr_gain
        \centering
        
        % Internal command of the figure for the automatic sizing
        % Path of the files
        \newcommand{\PathFig}{Fig_CL_corr_gain/}
        
	\newcommand{\subfigColor}{black}        
        
        \subfigimg[width=0.85\linewidth,pos=ul,font=\fontfig{\subfigColor}]{}{0.0}{\PathFig gain_curv}
              
        \caption{\label{fig:CL_corr_gain} Sensitivity of the closed loop estimator for different numbers of controlled modes in the noise limited regime (coloured curves). The coloured areas emphasise the~$\pm3\,\sigma$ regions. The black dashed (resp. plain) line is the fitted (resp. theoretical) sensitivity of~$\gcl \simeq 0.7$ (resp. 1). The coloured dashed lines emphasise the lateral error at which the panels of \reffig{fig:CL_corr_gain_sanity_check} are displayed.
        }
\end{figure}

The sensitivity of the closed loop estimator was assessed by injecting known lateral errors~$\Vdelta^\Tag{th}$ in the system, with amplitudes ranging from~$\Abs{\Vdelta^\Tag{th}}=\percent{0}$ to \percent{70} of a subaperture pitch~$\SApitch$ every~\percent{5} and angles ranging from~$\Arg{\Vdelta^\Tag{th}}=\SI{0}{\degree}$ to \SI{355}{\degree} every~\SI{5}{\degree}. For each case, batches of 500 consecutive closed loop iterations in a pure noise regime wihtout any turbulence were gathered for different numbers of controlled modes. The estimated lateral errors~$\tilde{\Vdelta}$ were de-rotated with the known angles~$\Arg{\Vdelta^\Tag{th}}$ to compute its mean value and its standard deviation for each amplitude~$\Abs{\Vdelta^\Tag{th}}$. The results are shown in \reffigs{fig:CL_corr_gain}{fig:CL_corr_gain_sanity_check}.

From \reffig{fig:CL_corr_gain}, it appears that the estimator follows a linear law for errors lower than \percent{25} of~$\SApitch$ for $\nmodes\leq800$. As expected, the more modes are used and the less noisy is the estimator. Nonetheless, the linearity range decreases with the number of modes, suggesting limitation in the approximation of \refeq{eq:corr_0}. Indeed, estimating the lateral error via \refeq{eq:CL_estim} does not solve the inverse problem of \refeq{eq:corr_theta} but its Taylor expansion of \refeq{eq:corr_00} for negligible coefficients~$\theta \sim 0$. As the correlation curves depend on this coefficient, see \reffig{fig:CL_corr_curves}, this can bias the estimator. The saturation of the estimator and the associated increase of its standard deviation  for~$\nmodes\geq500$ and for $\Abs{\Vdelta^\Tag{th}}>\percent{50}$ are linked with the limit of stability of the loop at large mis-registrations. This leads to DM command clipping and thus non-linearities in the system. This limit is reached at a smaller lateral error of $\Abs{\Vdelta^\Tag{th}}=\percent{20}$ when controlling an extreme number of modes of~$\nmodes = 1200$.

Surprisingly, the sensitivity of the estimator, fitted for~$\nmodes=250$, converges for small lateral errors towards:
\begin{equation}
        \label{eq:gain_cl}
        \gcl \simeq 0.70 < 1
        % 0.710547
        \,.
\end{equation}
This small value of $\gcl$ cannot only be explained by the approximation of \refeq{eq:corr_0}. It could come from other approximations made in \refsec{sec:method_CL_estim}. First, all the theory was derived for pure (and thus infinite) spatial frequencies. But the system is size-limited and with a circular pupil. There is not a strict definition of symmetric and anti-symmetric spatial frequency modes. This could lead to some coupling across spatial frequencies not accounted for by the model. Such coupling could also be induced in the command matrix, obtained by filtering the PSIM with~$\nmodes$ KL modes. In addition, estimating the correlation via~\refeq{eq:corr_empirical} assumes that the expectancy of the ratio is equivalent to the ratio of the expectancies of~\refeq{eq:corr}:
\begin{equation}
    % \corrt{cl} \sim \Avg{\frac{\cf{1}\cfconj{2}}{\sqrt{\cf{1}\cfconj{1}}\sqrt{\cf{2}\cfconj{2}}}}
    % \neq \frac{\Avg{\cf{1}\cfconj{2}}}{\sqrt{\Avg{\cf{1}\cfconj{1}}\Avg{\cf{2}\cfconj{2}}}}
    \corrt{cl} \sim \Avg{\frac{\Vc_{1}\conj{\Vc}_{2}}{\sqrt{\Vc_{1}\conj{\Vc}_{1}}\sqrt{\Vc_{2}\conj{\Vc}_{2}}}}
    \neq \frac{\Avg{\Vc_{1}\conj{\Vc}_{2}}}{\sqrt{\Avg{\Vc_{1}\conj{\Vc}_{1}}\Avg{\Vc_{2}\conj{\Vc}_{2}}}}
    \,.
\end{equation}
Further understanding the value of this sensitivity was not considered critical as any additional uncorrelated noise in the empirical estimator of \refeq{eq:corr_empirical} would lead to an underestimation of the correlation, via the normalisation by the variances.

Thus, the closed loop estimator is not absolute but relative, sensitive to the presence of a lateral error. It is thus relevant in the context of a corrective loop to act on the lateral error, whether mechanically (via a physical actuator as in GPAO) or numerically (via an update of the synthetic model of the instrument and its command matrix until convergence is achieved). The sensitivity is known to be smaller than one, insuring a stable convergence.

\reffigfull{fig:CL_corr_gain_sanity_check} shows the best fit correlation curves:
\begin{align}
    \corrt{t}\Paren{f} 
    & {}\triangleq{} \argmin{\corr{}}{\sum_{\Vk\in\kctrl}\Paren{\corrt{cl}\Paren{\Vk,f} - 2\pi\corr{\,}\Vk\scaprod\tilde{\Vdelta}}^2}
    \\
    \label{eq:corr_t}
    & {}={} \frac{\sum_{\Vk\in\kctrl}\corrt{cl}\Paren{\Vk,f}\Vk\scaprod\tilde{\Vdelta}}{\sum_{\Vk\in\kctrl} 2\pi\Paren{\Vk\scaprod\tilde{\Vdelta}}^2}
    \,,
\end{align}
in \refsubfigs{fig:CL_corr_gain_sanity_check}{a,c,e,g}  as well as the maps of the best fit correlation coefficients:
\begin{align}
    \corrt{2D}\Paren{\Vk} 
    & {}\triangleq{} \argmin{\corr{}}{\sum_{f>0}\Paren{\corrt{cl}\Paren{\Vk,f} - \Vcorr{0}\Paren{f}\corr{}}^2}
    \\
    \label{eq:corr_2D}
    & {}={} \frac{\sum_{f>0}\corrt{cl}\Paren{\Vk,f}\Vcorr{0}\Paren{f}}{\sum_{f>0} \Vcorr{0}^2\Paren{f}}
    \,,
\end{align}
in \refsubfigs{fig:CL_corr_gain_sanity_check}{b,d,f,h} for some specific~$\Vdelta^\Tag{th}$. These curves and maps are not used to compute the estimated misalignments $\tilde{\Vdelta}$ which are obtained using \refeq{eq:CL_estim}. Nonetheless, they are good indicators to validate our AO loop modelling and its assumptions.

\begin{figure}[t!] % fig:CL_corr_gain_sanity_check
    \centering
    
% Internal command of the figure for the automatic sizing
    % Path of the files
    \newcommand{\PathFig}{Fig_CL_corr_gain/}
    
    \newcommand{\subfigColor}{black}
    
    % Line ratio
    \newcommand{\LineRatio}{0.985}
    
    % Vertical space between lines
    \newcommand{\spaceLine}{-0.05cm}

	% Font of the text in the figure
%	\newcommand{\fontTxt}[1]{\textbf{\small #1}}
	\newcommand{\fontTxt}[1]{\scriptsize{#1}}
        
	% Width of the text boxes        
	\newcommand{\widthTxt}{11pt}
        
	% Width of the text boxes        
	\newcommand{\sizeTxt}[1]{\Large{#1}}    
    
    \newcommand{\widthFig}{\dimexpr (\linewidth - \widthTxt)}
    
    \newcommand{\suffig}{_250_angle_\thetaGain_shift_15}
    
    \newcommand{\figOne}[1]{\PathFig corr_curv_modes#1}
    \newcommand{\figTwo}[1]{\PathFig corr_map_modes#1}
    
    % Getting the size of the boxes
    \sbox1{\includegraphics{\figOne{\suffig}}}
    \sbox2{\includegraphics{\figTwo{\suffig}}}
    
    % Defining column width command
    
	\newcommand{\ColumnWidth}[1]
		{\dimexpr \LineRatio \widthFig * \AspectRatio{#1} / (\AspectRatio{1} + \AspectRatio{2}) \relax
		}
	\newcommand{\ColumnGap}{\hspace {\dimexpr \widthFig /2 - \LineRatio\widthFig /2 }}    
    
    % Figure table
    \begin{tabular}{
        @{}
		M{\widthTxt}
		@{}
        M{\ColumnWidth{1}}
        @{\ColumnGap}
        M{\ColumnWidth{2}}
        @{}
        }
        
		\rotatebox[origin=l]{90}{\fontTxt{$\nmodes=250$ / $\Abs{\Vdelta^\Tag{th}}=\percent{15}\SApitch$}}
		&
        \subfigimg[width=\linewidth,pos=ur,font=\fontfig{\subfigColor}]{$\,$(a)}{0.0}{\figOne{\suffig}} &
        \subfigimg[width=\linewidth,pos=ur,font=\fontfig{\subfigColor}]{$\,$(b)}{0.0}{\figTwo{\suffig}}
        \\[\spaceLine]
		\rotatebox[origin=l]{90}{\fontTxt{$\nmodes=500$ / $\Abs{\Vdelta^\Tag{th}}=\percent{45}\SApitch$}}
		&
        \subfigimg[width=\linewidth,pos=ur,font=\fontfig{\subfigColor}]{$\,$(c)}{0.0}{\figOne{_500_angle_\thetaGain_shift_45}} &
        \subfigimg[width=\linewidth,pos=ur,font=\fontfig{\subfigColor}]{$\,$(d)}{0.0}{\figTwo{_500_angle_\thetaGain_shift_45}}
        \\[\spaceLine] 
		\rotatebox[origin=l]{90}{\fontTxt{$\nmodes=800$ / $\Abs{\Vdelta^\Tag{th}}=\percent{65}\SApitch$}}
		&
        \subfigimg[width=\linewidth,pos=ur,font=\fontfig{\subfigColor}]{$\,$(e)}{0.0}{\figOne{_800_angle_\thetaGain_shift_65}} &
        \subfigimg[width=\linewidth,pos=ur,font=\fontfig{\subfigColor}]{$\,$(f)}{0.0}{\figTwo{_800_angle_\thetaGain_shift_65}}
        \\[\spaceLine]
		\rotatebox[origin=l]{90}{\fontTxt{$\nmodes=1200$ / $\Abs{\Vdelta^\Tag{th}}=\percent{10}\SApitch$}}
        &
        \subfigimg[width=\linewidth,pos=ur,font=\fontfig{\subfigColor}]{$\,$(g)}{0.0}{\figOne{_1200_angle_\thetaGain_shift_10}} &
        \subfigimg[width=\linewidth,pos=ur,font=\fontfig{\subfigColor}]{$\,$(h)}{0.0}{\figTwo{_1200_angle_\thetaGain_shift_10}}
        \\[-0.15cm]
        &
        $\;\;\;$Frequency $f$ (\SI{}{\hertz})
        &
        Spatial frequency $\Vk$ %(\SI{}{\per\meter})
    \end{tabular}  
          
    \caption{\label{fig:CL_corr_gain_sanity_check} Sanity check of the fits introduced in \reffig{fig:CL_corr_gain} (arbitrary unit) for different numbers of controlled modes~$\nmodes$ and shift amplitudes $\Abs{\Vdelta^\Tag{th}}$ ($\Arg{\Vdelta^\Tag{th}}=\SI{\thetaGain}{\degree}$). \refpans{a,c,e,g}: curves of the best temporal correlation fit~$\Vcorrt{t}$ (grey). Black curves: theoretical correlation curve~$\Vcorr{0}$ for~$\Vdelta\rightarrow 0$. Coloured curves: mean filter of~$\Vcorrt{t}$ with a sliding window of $\pm5$ data points. \refpans{b,d,f,h}: maps of the best fit correlation coefficients~$\Vcorrt{2D}$.
    }
\end{figure}

From the curves of \refsubfigs{fig:CL_corr_gain_sanity_check}{a,c,e,g}, it first appears that the temporal correlation of the commands exhibits the expected signal and with the correct shape (black curve), validating the formalism of \refsec{sec:method_CL_estim}. As expected, this signal is noisier for small lateral errors, see \refsubfigs{fig:CL_corr_gain_sanity_check}{a,g} versus \refsubfig{fig:CL_corr_gain_sanity_check}{b}. For increased mis-registration parameters, divergences from the approximation of \refeq{eq:corr_0} start to appear, see \refsubfigs{fig:CL_corr_gain_sanity_check}{b,c}. The correlation peak slightly shifts towards lower temporal frequencies, as predicted by the curves of \reffig{fig:CL_corr_curves} for large coefficients~$\theta$. In addition, beyond the limit of loop stability, as in \refsubfig{fig:CL_corr_gain_sanity_check}{c}, the correlation curve presents artefacts related to the non-linearities induced by the command clipping. This nonetheless does not prevent the estimator from correctly obtaining the order of magnitude of the injected lateral error and its global orientation.

Then, the 2D maps of \refsubfigs{fig:CL_corr_gain_sanity_check}{b,d,f,h} markedly show the `tip-tilt' feature expected along the mis-registration direction from \refeq{eq:theta}. This feature is restricted to a disk, the radius of which depends on the number modes, validating \refeq{eq:kctrl}. As for the temporal correlation, the noise level on the 2D maps is higher for small lateral errors, see \refsubfigs{fig:CL_corr_gain_sanity_check}{b,h} versus \refsubfigs{fig:CL_corr_gain_sanity_check}{d,f}. Once again, artefacts are visible beyond the limit of loop stability, as in \refsubfig{fig:CL_corr_gain_sanity_check}{c}:  on the highest controlled frequencies, the `tip-tilt' feature is broken.

As a final remark, all these results were obtained using batches of only 500 telemetry frames. This is significantly lower than previous methods based on telemetry \citep{Bechet:12_SPIE, Kolb:12_SPIE_AOF} for which tens of thousand frames must be acquired to first estimate the interaction matrix from the measurements. The proposed method does not need this demanding intermediate step and efficiently uses the available signal in the Fourier domain where the mis-registration signature is sparse, optimising the signal over noise ratio. We notice here that the impact of the number of frames in the telemetry batches is discussed in \refapp{app:batch_size}.

\subsubsection{Bias induced by a frozen flow turbulence}

\label{sec:CL_estim_sim_wind}

In this section, turbulence is added to the system. Indeed, as mentioned by \citet[in Sect. 4.3]{Heritier:19_PHD} and discussed in \refapp{app:turb_prop}, the wind could bias the closed loop estimator when the temporal error is the dominant source of perturbation. To emulate the turbulence, a phase screen was generated for each simulation using the power spectrum density method proposed by~\citet{McGlamery:76_Turb_PSD} on a domain twelve times larger than the telescope pupil. To keep the phase screen 2D-periodic, no sub-harmonic was added~\citep{Lane:92_Subharmonics}. The screen was looped when translating the frozen flow according to the wind speed~$\vWind$ and direction~$\thetaWind$. All simulations were done for a typical Fried parameter of~$\rFried = \SI{12}{\centi\meter}$ at~$\lambda_{0}=\SI{500}{\nano\meter}$.

\begin{figure}[t!]
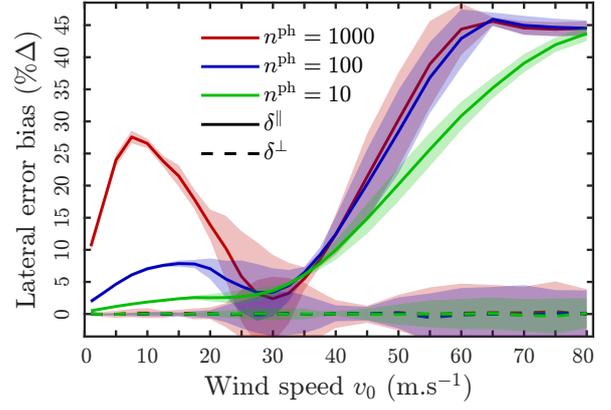
 % fig:CL_corr_wind_bias
    \centering
    
	% Internal command of the figure for the automatic sizing
    % Path of the files
    \newcommand{\PathFig}{Fig_CL_corr_wind_bias/}
    
    \newcommand{\subfigColor}{black}        
    
    \subfigimg[width=0.85\linewidth,pos=ul,font=\fontfig{\subfigColor}]{}{0.0}{\PathFig wind_bias}
          
    \caption{\label{fig:CL_corr_wind_bias} Influence of the wind speed~$\vWind$ and number of photons~$\nph$ on the estimation of the lateral error~$\Vdelta = \Paren{\delta^\parallel, \delta^\perp}$ in the wind frame, parallel (plain curves) and perpendicular (dashed curves) to the wind direction~$\thetaWind$. The coloured areas emphasise the~$\pm1\,\sigma$ regions.}
\end{figure}

To assess the impact of a frozen flow turbulence on the closed loop estimator, both in terms of strength and orientation, simulations were run for wind speeds ranging from~$\vWind = \SI{1}{\meter\per\second}$ to \SI{80}{\meter\per\second}. The simulations being time-consuming, we used the symmetry of the problem to test wind directions only ranging from~$\thetaWind = \SI{0}{\degree}$ to \SI{90}{\degree} every~\SI{7.5}{\degree}. For each simulation, the system was initialised without any error~$\Vdelta=\V{0}$. $\nmodes=500$ modes are controlled. Batches of 500 consecutive closed loop iterations (\SI{0.5}{\second}) are gathered to feed the closed loop estimator. In between each batch, the lateral error in the system was updated as follows:
\begin{equation}
    \label{eq:corrective_loop}
    \Vdelta^{i+1} = \Vdelta^{i} + \Thong{}\, \tilde{\Vdelta}
    \,,
\end{equation}
with the gain~$\Thong = 0.5$. Combined with the closed loop estimator sensitivity~$\gcl$, \refeq{eq:gain_cl}, this leads to an approximated gain of $0.35$ that insures a stable and slow convergence. The convergence values were obtained as described in \refapp{app:wind_conv}. Results are gathered in \reffig{fig:CL_corr_wind_bias} for different photon noise regimes and for both the bias parallel (plain curves) and perpendicular (dashed curves) to the wind direction.

It first appears that as expected, the induced bias is along the wind direction (plain vs dashed curves). As seen by \citet[Sect. 4.3]{Heritier:19_PHD} with another mis-registration estimator, the bias amplitude depends on the ratio between the noise and the temporal error as well as the wind speed. For standard photon noise levels of $\nph$ between 10 and 100, it stays below \percent{10} of a subaperture for wind speed below \SI{37.5}{\meter\per\second}, that is to say below \SI{20}{\milli\meter}. This should be compared to the total VLT pupil size of $D=\SI{8}{\meter}$.

Close to the noiseless regime, with $\nph = 1000$, the bias peaks a bit under \percent{30} of a subaperture for low wind speeds. Some simulations, not shown here, indicate that this peak depends on the simulation resolution, suggesting an aliasing problem in the interpolation when translating the frozen screen. %This could also explain why the dispersion on the bias increases for wind speeds higher than \SI{25}{\meter\per\second}.
Better characterising this peak was not considered critical as such a situation would not be met in a real AO system. Indeed, errors which are not considered here would add up. Among others: the sensor readout noise and saturation, and some AO system non linearities such as the spot centroiding method or DM command clipping or saturation. And as discussed in \refapp{app:turb_prop}, the mis-registration estimator works better in a noise-limited regime. Considering only the photon noise is thus a conservative hypothesis.

Things worsen for wind speeds higher than $\vWind\geq\SI{37.5}{\meter\per\second}$. For all the noise regimes, the estimator bias increases with the wind speed, until reaching a plateau around~$\Abs{\Vdelta}=\percent{45}\SApitch$. This saturation corresponds to the limit at which the AO loop becomes unstable when controlling 500 modes. This instability overcomes the wind perturbation in the loop global disturbance, stabilising the estimator. This result must be interpreted carefully. Indeed, for such high wind speeds, our end-to-end simulations reach their limits: during the exposure time, the wind evolution is not negligible and it should be averaged by simulating subframe steps.

As a consequence, the closed loop estimator was further tested on a more realistic turbulence, defined as the reference in the GPAO requirements \citep[Sect. 6]{LeBouquin:23_GPAO_design}. \reftab{tab:GPAO_requirements} gathers the parameters of its 5 layers. Its Fried parameter is~$\rFried=\SI{10}{\centi\meter}$. This corresponds to a seeing of $\SI{1}{\arcsecond}$. This system is initialised with a lateral misalignment of $\Abs{\Vdelta}=\percent{70}\SApitch$ and $\Arg{\Vdelta}=\SI{35}{\degree}$, corrected every batches of 500 telemetry commands via \refeq{eq:corrective_loop} and still controlling $\nmodes=500$ modes.

\begin{table}[t!] % tab:GPAO_requirements
    \caption{\label{tab:GPAO_requirements} Atmosphere layer parameters of the GPAO requirements.}
    \centering
    \begin{tabular}{cccccc}
    \hline
    \hline
    Layer & 1 & 2 & 3 & 4 & 5 
    \\
    \hline
    $C_n^2$ & 0.67 & 0.07 & 0.1 & 0.1 & 0.06
    \\
    $\vWind$ & 12.2 & 8.3 & 30.3 & 56 & 32.5
    \\
    $\thetaWind$ & 150.1 & 79.6 & -70 & -7.7 & -82.6
    \\
    \hline
    \end{tabular}
    %\tablefoot{Coucou}
\end{table}

\begin{figure}[t!]
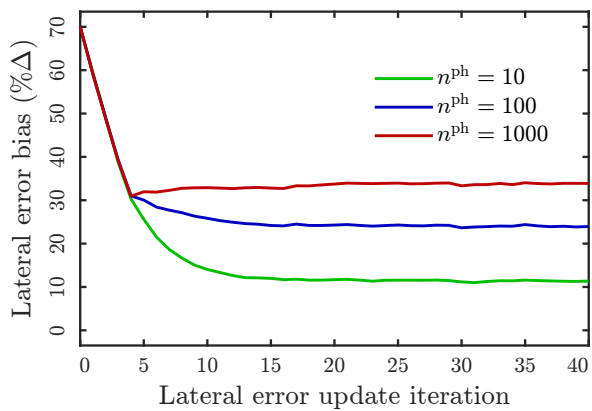
 % fig:CL_corr_GPAO_requirements

    \centering
    
    % Internal command of the figure for the automatic sizing
    % Path of the files
    \newcommand{\PathFig}{Fig_CL_corr_GPAO_requirements/}
    
    \newcommand{\subfigColor}{black}        
    
    \subfigimg[width=0.85\linewidth,pos=ul,font=\fontfig{\subfigColor}]{}{0.0}{\PathFig wind_bias}
          
    \caption{\label{fig:CL_corr_GPAO_requirements} Convergence of the lateral error~$\Abs{\Vdelta}$ in function of the update iteration for the turbulence case of the GPAO requirements, see \reftab{tab:GPAO_requirements}, and for different photon noise levels~$\nph$.}
\end{figure}

The evolution of the lateral errors with the update iteration for different noise regimes is given in \reffig{fig:CL_corr_GPAO_requirements}. For the first four iterations, the system is unstable and the convergence curves overlap for all noise regimes. They quickly drop around~$\Abs{\Vdelta}=\percent{30}\SApitch$. Despite a layer with a wind speed beyond~$\SI{50}{\meter\per\second}$, the divergence of the estimator remains under control. For the lowest noise level (red), it stays in the stability regime under $\Abs{\Vdelta}=\percent{35}\SApitch$. For a realistic case of a noise dominated system (green), the corrective loop successfully tackles the initial lateral error and converges with a bias of~$\Abs{\Vdelta}\simeq\percent{10}\SApitch$.

We also emphasise here that real turbulence would be more complex than additive frozen flow layers. Any boiling or uncorrelated evolution would inject extra noise in the system producing the expected correlation curves, as explained in the first part of \refapp{app:turb_prop}. This would further participate to make the estimator robust to the frozen flow parts of the wind.

\subsection{Experimental results: Auto-alignment in operation}
\label{sec:exp_CL_estim}

In this section, the AO loop was closed using the parameters listed in \reftab{tab:list_variable}. To test the closed loop estimator, we introduced a random lateral error to place the system at the limit of stability with the phase plate spinning at its maximal speed ($\vWind\simeq\SI{8.4}{\meter\per\second}$, ~$\rFried \simeq \SI{14}{\centi\meter}$). This configuration leads to an unstable wave pattern on the DM commands (see \refsubfig{fig:CL_corr_bench}{c}), which is seen by the WFS (see \refsubfigs{fig:CL_corr_bench}{a,b}) and that is typical to a lateral mis-registration \citep{Heritier:18_SPIE_PyWFS_fitting}. In addition, we kept the angle of \SI{35}{\degree} between the SH-WFS pupil and the DM grid. This angle can clearly be seen in the pattern orientation in between \refsubfigs{fig:CL_corr_bench}{a,b} and \refsubfig{fig:CL_corr_bench}{c}.

\begin{figure}[t!] % fig:CL_corr_bench
    \centering
    
	% Internal command of the figure for the automatic sizing
    % Path of the files
    \newcommand{\PathFig}{Fig_CL_corr_bench/}

    % Vertical space between lines
    \newcommand{\spaceLine}{-0.05cm}
    
    % Font of the text in the figure
    \newcommand{\fontTxt}[1]{\textbf{\tiny #1}}
    
    % Line ratio
    \newcommand{\LineRatio}{0.99}
    % Color of the panel letter
    \newcommand{\subfigColor}{white}
    % Defining column width command
    \newcommand{\ColumnWidth}
        {\dimexpr \LineRatio \linewidth /3 \relax
        }
    \newcommand{\ColumnGap}{\hspace {\dimexpr \linewidth /2 - \LineRatio\linewidth /2 }}
    % Figure table
    \begin{tabular}{
        @{}
        M{\ColumnWidth}
        @{\ColumnGap}
        M{\ColumnWidth}
        @{\ColumnGap}
        M{\ColumnWidth}
        @{}
        }
        
        % Title line    
        \fontTxt{WFS pixels}
        &
        \fontTxt{WFS $y$-slopes}
        &
        \fontTxt{DM command}
        \\             
        \subfigimg[width=\linewidth,pos=ul,font=\fontfig{\subfigColor}]{$\,$(a)}{0.0}{\PathFig Pixel} &
        \subfigimg[width=\linewidth,pos=ul,font=\fontfig{\subfigColor}]{$\,$(b)}{0.0}{\PathFig Slopes} &
        \subfigimg[width=\linewidth,pos=ul,font=\fontfig{\subfigColor}]{$\,$(c)}{0.0}{\PathFig DM}
    \end{tabular}
    
    \vspace{\spaceLine}

    \renewcommand{\subfigColor}{black}
    \renewcommand{\LineRatio}{0.985}
    \newcommand{\figOne}{\PathFig corr_curv}
    \newcommand{\figTwo}{\PathFig corr_map}
    
    % Getting the size of the boxes
    \sbox1{\includegraphics{\figOne}}
    \sbox2{\includegraphics{\figTwo}}
    
    % Defining column width command
    \renewcommand{\ColumnWidth}[1]
        {\dimexpr \LineRatio \linewidth * \AspectRatio{#1} / (\AspectRatio{1} + \AspectRatio{2}) \relax
        }
    \renewcommand{\ColumnGap}{\hspace {\dimexpr \linewidth /1 - \LineRatio\linewidth /1 }}
    
    % Figure table
    \begin{tabular}{
        @{}
        M{\ColumnWidth{1}}
        @{\ColumnGap}
        M{\ColumnWidth{2}}
        @{}
        }
        
        \subfigimg[width=\linewidth,pos=ur,font=\fontfig{\subfigColor}]{$\,$(d)}{0.0}{\figOne} &
        \subfigimg[width=\linewidth,pos=ur,font=\fontfig{\subfigColor}]{$\,$(e)}{0.0}{\figTwo}
        \\[-0.15cm]
        $\;\;\;$Frequency $f$ (\SI{}{\hertz})
        &
        Spatial frequency $\Vk$ %(\SI{}{\per\meter})
    \end{tabular}  
    
    \caption{\label{fig:CL_corr_bench} Example in the GPAO bench of the closed loop estimator. \refpans{a,b,c}: SH-WFS pixel frame, WFS $y$-slopes and DM commands at the limit of stability before the alignment. Green circles: WFS pupil edges. \refpan{d}: curves of the best temporal correlation fit~$\Vcorrt{t}$ (grey), of the mean filter of~$\Vcorrt{t}$ with a sliding window of $\pm5$ data points (green), and of the theoretical correlation~$\Vcorr{0}$ (black). \refpan{e}: map of the best fit correlation coefficients~$\Vcorrt{2D}$. The green circle encompasses the frequency space theoretically controlled by the command matrix.}
\end{figure}

We recall here that, depending on the context, the estimator can be used in two different ways: (1) to iteratively update the parameters of the command matrix of the system with a fixed lateral error until convergence or (2) to dynamically correct the system alignment until nominal registration is reached. The needs of GPAO introduced in \refsec{sec:intro} correspond to this latter option. As for the simulations of \refsec{sec:CL_estim_sim_wind}, the gain of the corrective secondary loop in charge of realigning the system is set to 0.5. Combined with the closed loop estimator sensitivity~$\gcl$, \refeq{eq:gain_cl}, this leads to an approximated gain of $0.35$.

\begin{figure}[t!]
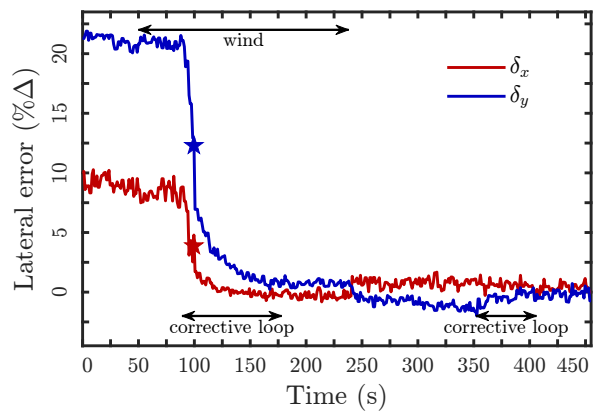
 % fig:CL_corr_bench_conv
    \centering
    
    % Internal command of the figure for the automatic sizing
    % Path of the files
    \newcommand{\PathFig}{Fig_CL_corr_bench/}
   
    \subfigimg[width=0.85\linewidth,pos=ul,font=\fontfig{black}]{}{0.0}{\PathFig Conv_curv}
    
    \caption{\label{fig:CL_corr_bench_conv}
    Convergence of the lateral error corrective loop with the closed loop estimator. Double arrows indicate when the phase plate and the corrective loop are active. The stars indicate the iteration at which \refsubfigs{fig:CL_corr_bench}{d,e} are displayed.}
\end{figure}

As highlighted on \reffig{fig:CL_corr_bench_conv}, the corrective loop is run in two different situations. First, the phase plate is spinning. Then the corrective loop is opened and the phase plate is stopped to assess the bias induced by the wind on the closed loop estimator. Finally, a new set of corrective loop iterations is performed. In between all these steps, an IM is measured (with the phase plate stopped) to quantitatively retrieve the system misalignment with the AOF tool in SPARTA. The results are gathered in \reftab{tab:CL_corr_bench}.

\begin{table}[t!] % tab:CL_corr_bench
    \caption{\label{tab:CL_corr_bench} System auto-alignment with the closed loop estimator}
    \centering
    \begin{tabular}{cccc}
    \hline
    \hline
     &  $\delta_x~/~\delta_y$ & $\theta$ & $\delta_{\rho_x}~/~\delta_{\rho_y}$
    \\& ($\%\SApitch$) & (\SI{}{\degree}) & ($\%$)
    \\
    \hline
    $t=\SI{0}{\second}$ & 11.3~/~44.7 & 35.1 & 0.2~/~1.2
    \\
    $t=\SI{300}{\second}$ & -1.4~/~-2.25 & 35.1 & 0.3~/~1.1
    \\
    $t=\SI{450}{\second}$ & -1.9~/~-1.6 & 35.1 & 0.2~/~1.0
    \\   
    \hline
    \end{tabular}
    \tablefoot{PSIM parameters fitted by SPARTA on the IM measured before the system auto-alignment with the closed loop estimator ($t=\SI{0}{\second}$), after the convergence while the phase plate was spinning ($t=\SI{300}{\second}$) and without the phase plate ($t=\SI{400}{\second}$).  $\theta$: clocking. $\delta_{\rho_x}~/~\delta_{\rho_y}$: stretches along the $x$ and $y$-axes.}
\end{table}

% ----- START -----
% Variable / 2D_corr / SPARTA
% Amplitude: 2.405221e+00 / 2.697546e+00
% xShift: 9 / 1.135341e+01
% yShift: 51 / 4.466575e+01
% Rotation: 3.499878e+01 / 3.506494e+01
% xStretch: 0 / 1.568440e-01
% yStretch: 0 / 1.163002e+00
% ----- START -----

% ----- CONV1 -----
% Variable / 2D_corr / SPARTA
% Amplitude: 4.369899e+00 / 4.817702e+00
% xShift: 18 / 1.518117e+01
% yShift: -27 / -2.329552e+01
% Rotation: 3.499878e+01 / 3.507900e+01
% xStretch: 0 / 2.897277e-01
% yStretch: 0 / 1.140643e+00
% ----- CONV1 -----

% ----- CONV2 -----
% Variable / 2D_corr / SPARTA
% Amplitude: 4.365340e+00 / 4.821910e+00
% xShift: 17 / 1.464708e+01
% yShift: -26 / -2.262946e+01
% Rotation: 3.499878e+01 / 3.507802e+01
% xStretch: 0 / 3.037195e-01
% yStretch: 0 / 1.152877e+00
% ----- CONV2 -----

% ----- CONV2_OL -----
% Variable / 2D_corr / SPARTA
% Amplitude: 2.441844e+00 / 2.738880e+00
% xShift: -4 / -1.948066e+00
% yShift: 1 / -1.587420e+00
% Rotation: 3.499878e+01 / 3.508555e+01
% xStretch: 0 / 2.054324e-01
% yStretch: 0 / 1.038362e+00
% ----- CONV2_OL -----

Despite starting in an unstable state, the corrective secondary loop efficiently converges in a few tens of iterations. At the start and stop of the phase plate, a small change in $\tilde{\Vdelta}$ can be seen \reffig{fig:CL_corr_bench_conv}, suggesting that the frozen wind indeed induces a bias in the estimator. We further insist here that the values given in the figure are the lateral error computed by the closed loop estimator for each acquired telemetry batch of 500 frames. They consequently suffer from the underestimated sensitivity discussed in \refsec{sec:CL_estim_sim_gain}. For example, at the initial state, $\tilde{\Vdelta}\simeq\Paren{8,21}\%\SApitch$ but SPARTA indicates that the real error is closer to $\tilde{\Vdelta}\simeq\Paren{11,45}\%\SApitch$, see \reftab{tab:CL_corr_bench}.

With or without the phase plate, the convergence is stable but noisy with oscillations of a few percent. The IM measured after the corrective loop convergence with the phase plate spinning gives that $\Abs{\Vdelta}\simeq\percent{2.7}\SApitch$. This bias is lower than the predictions of the simulations introduced in \refsec{sec:CL_estim_sim_wind} suggesting that they were indeed conservative. Without the phase plate spinning, the lateral errors slightly drops to $\Abs{\Vdelta}\simeq\percent{2.5}\SApitch$ which is in the noise level of the estimator as shown in \reffig{fig:CL_corr_bench_conv}.

For information and sanity check, \refsubfigs{fig:CL_corr_bench}{d,e} give the correlation curve~$\Vcorrt{t}$ and map~$\Vcorrt{2D}$ for the point indicated with a star in \reffig{fig:CL_corr_bench_conv}. Looking at \refsubfig{fig:CL_corr_bench}{d}, the main feature is a mismatch between the empirical correlation (grey/green) and the theoretical curve (black) at small frequencies. This discrepancy can be attributed to the wind, as discussed in \refapp{app:turb_prop} and seen in frozen flow simulations.

Finally, the map of \refsubfig{fig:CL_corr_bench}{e} shows that the command correlation leaves a strong trace in the 2D spatial frequency space with the typical `tip-tilt' pattern. This pattern matches the theoretical area of the $\nmodes=500$ delimited by the coloured circle.

As a conclusion, these results prove that the closed loop estimator is able to correctly monitor and feed the corrective secondary loop. They also show that the theoretical model can successfully fit real data. In addition, in these tests, the wind induced by the phase plate had a negligible impact.

\section{Conclusions and perspectives}

In this work, we introduce two novel methods to estimate the lateral misalignment of an AO system. First, their underlying theory is presented. Then, we describe how their performance was tested in simulations. Finally, they were experimentally validated in the GPAO development bench with a system still largely under development and not yet fully characterised. These methods will be at the core of the GPAO system auto-alignment strategy while in operation to keep the AO performances at their optimal level throughout the observations. These methods were presented in the context of the GPAO system for its \X{40} SH-WFS mode. They were also successfully tested in the bench for the other GPAO configurations (\X{30} and \X{9}) and, as discussed below, they can be adapted to other AO systems, in particular for the future ELTs.

\subsection{Perturbative modal estimator}
\label{sec:conclu_2D_corr_estim}

This estimator uses the spatial 2D representation of the interaction matrix to look for spatial correlations of the measures with a reference. Not based on a complex model fitting, this method has the advantages to be fast and reliable, with a large capture range and a high S/N and is robust to loosely constrained higher-order mis-registrations such as clocking (to a few degrees) or stretch (to a few percent). An interesting by-product of the method is an estimation of the IM amplitude.

In this work, the reference is a synthetic IM, but an experimental reference measured during calibration could also be used, avoiding the need to rely on a possibly complex PSIM model of the system. A possible evolution of this method could be to totally get rid of the need of a reference IM. For example, by focussing on purely symmetric or anti-symmetric modes, looking at the symmetries of the modes in the measured IM and their auto-correlation should suffice to estimate the system decentering.

In this paper, we focus on a pertubative approach in open loop where the modal IM is actively measured by injecting modes in the system. We also tested that the method works in closed loop. The results are not presented in this paper but the lateral error is successfully retrieved. Naturally, the amplitude estimate is not reliable as the perturbations are fought by the AO loop. 

In this work, we assume that the modal IMs are obtained with an invasive approach such as fast `push-pull' of the modes \citep{Kasper:04_Hadamard, Oberti:06_SPIE_PSIM_vs_measured_IM, Kolb:16_SPIE_Review_AO_calibration} or via command modulation \citep{Esposito:06_SPIE_modulation, Pinna:12_SPIE_FLAO, Kolb:16_SPIE_Review_AO_calibration}. The next step will be to test the estimator on the IM fully estimated from loop telemetry, for example as described by \citet{Bechet:12_SPIE}, based on the correlation between the slopes and the commands. Such IMs are generally noisy, but we have shown in this study that the estimator is robust to a high level of noise. Doing so would make the method non-pertubative, but potentially biased by the wind in case of a strong frozen flow, as discussed by \citet[in Sect. 4.3]{Heritier:19_PHD}.

The method could also be extended to measure higher orders of mis-registration such as clocking or stretches. Nonetheless, these problems do not have a simple analytical solution as  for the lateral errors, \refeq{eq:2D_corr_map} and would surely imply iterative fitting of the parameters or extensive traversal of the parameter space.

In this work, we studied the influence of the number of KL modes in \refapp{app:KL_modes}, but we did not dig into optimising the subset of modes in order to maximise the estimator performances. We chose the pragmatic approach to select modes that will be seen and controlled by all the GPAO modes. Nonetheless, it is known that different modes do not provide the same sensitivity to the mis-registration parameters \citep{Heritier:18_SPIE_PyWFS_fitting}. We could optimise the modal shapes to improve the lateral resolution or the capture range and to avoid angular redundancy to prevent multiple secondary maxima in individual modal correlations or if looking for the clocking for example. Nevertheless, \citet{Heritier:21_SPRINT} have shown that fine tuning the modes is not critical.

Finally, this method was presented in the context of a SH-WFS, but it could be extended to other spatial WFS. For example, it could be used for the sub-pixel alignment of the pupil images of a pyramid WFS which also contain spatial information.

\subsection{Non-perturbative closed loop estimator}

This estimator uses the temporal correlation of the noise in the 2D representation of the DM commands in the closed loop telemetry to identify cross-talks between symmetric and anti-symmetric spatial frequencies. The main advantage of this method, and at the origin of its development, is its sparsity in the Fourier domain. Thus, this method is fast and particularly suited for systems with ever increasing numbers of actuators such as future ELTs, both in terms of execution time and memory requirements. GPAO will serve as an on-sky demonstrator of this technique, which is the current baseline to monitor and drive the quaternary DM for the ESO ELT in its engineering single conjugated AO mode, based on a pyramid WFS operating in the near infrared \citep{Bonnet:23_AO4ELT_ELT}.

In addition, in the frequency domain, the impact of a frozen flow is also sparse, theoretically reducing the bias induced by the wind on the estimator. We nonetheless showed via simulations that for frozen flows with high wind speeds, the induced bias could diverge towards the loop stability limit. Nonetheless, the bias induced by the wind was not considered of high concern since the efficiency of the method interestingly increases with the noise in the system, as already intuited by \citet[Sect. 4.1.3]{Heritier:19_PHD}. This makes it suitable to AO loops pushed at their limits (low magnitude or limit of stability or linearity) and for real systems always noisier than simulations. Nevertheless, a possible evolution of the method would be to further work on its de-biasing, for example by setting the weights of the correlation curve fitting to zero for low temporal frequency corresponding to the maximal awaited wind-induced signal. As seen in \refsubfig{fig:CL_corr_wind_corr}{b}, this would depend on the spatial frequency insuring a correct number of data points through all the controlled space. Another solution would be to find a merit criterion on the correlation curves when solving \refeq{eq:CL_estim}. Indeed, the correlation signal induced by the wind strongly changes the shape of the theoretical correlation curves of the noise and should be quantifiable. Another direction to investigate is the exploitation of the real part of the correlation curves. It is discarded by the estimator of the proposed method, but the mixture of the noise propagation and the turbulence propagation should create a non-zero signal in the real part of the command correlation.

Another interesting features of the method is that it does not rely on a model of the AO system. Having access to the telemetry of a 2D spatial observable of the closed loop is sufficient. This could be the DM commands, as in this work, or directly the WFS measurements. This makes it particularly suitable to the pyramid WFS which directly gives the Hilbert transform of the incident wavefront perturbations in the telescope pupil for the non-modulated spatial frequencies. Nonetheless, these WFSs have a lower noise propagation than SH-WFSs, which could make them more sensitive to the wind bias, as noticed by \citet[Sect. 4.4]{Heritier:19_PHD}. Therefore, further quantitative analyses should be performed before translating the results to such WFSs.

In addition, this method depends on only a limited number of parameters. We have seen that the shape of correlation signals are very sensitive to these parameters. They were fixed in this study, but a possible evolution would be to fit these parameters along with the lateral errors in a global inverse problem approach. On top of potentially reducing the estimator bias, this could also provide a supplementary way to monitor the loop gain or RTC performances.

A main drawback of this closed loop estimator is that it does not provide absolute measurements. It is relative to the registration state used to generate the command matrix. We showed that the measurement sensitivity is a priori unknown but nonetheless less than one. This guarantees a stable convergence in closed loop. The method is therefore suitable both for optical compensation by means of a pupil steering actuator (as for GPAO) or numerical compensation by command matrix updates derived from the accumulated error measurements, as for example described by \citet{Heritier:21_SPRINT}.

Currently, this estimator was developed for lateral error monitoring which are `zero-order' mis-registrations. This means that they only couple symmetric and anti-symmetric parts of the same spatial frequencies. A work is undergoing to extend this method to `first-order' mis-registrations (clocking, magnification, and anamorphoses) with promising results. The difficult part is that such errors couple neighbouring spatial frequencies~$\Vk$ and~$\Vk+\delta\Vk$, increasing the complexity of the block diagram of \reffig{fig:CL_corr_spatial_coupling} and the resulting correlation curves. Nonetheless, the estimators remain sparse in the Fourier domain and thus fast and scalable to complex systems.

In conclusion, a long-term perspective for both estimators would be their adaptation beyond single conjugated AO to multi-conjugated AO systems composed of multiple DMs and multiple WFSs. If the adaptation of the pertubative 2D modal estimator were to be straightforward, this would be less so the case for the non-perturbative closed loop estimator. Indeed, mis-registrations between the different subsystems will produce numerous cross-talks potentially difficult to disentangle in the observable temporal correlations.

%-------------------------------------------------------------------%
%------------------------ Acknowledgements  ------------------------%
%-------------------------------------------------------------------%

\begin{acknowledgements}
    % This project has received funding from the European Un{\-}ion's Horizon 2020 research and innovation programme under grant agreement No 101004719.
    % This project has received funding from the European Research Council (ERC) under the European Union's Horizon 2020 research and innovation program under grant agreement CoG - 866070.
    This project has received funding from the European Un{\-}ion's Horizon 2020 research and innovation programme under grant agreements No 101004719 and European Research Council (ERC) / CoG - 866070.
\end{acknowledgements}

%-------------------------------------------------------------------%
%-------------------------- Bibliography  --------------------------%
%-------------------------------------------------------------------%

\bibliographystyle{aa}
\bibliography{bib_online_misreg_shift}

%\flushcolsend

%-------------------------------------------------------------------%
%--------------------------- Appendices  ---------------------------%
%-------------------------------------------------------------------%

\begin{appendix}

\section{Impact of the number modes on the 2D modal estimator}
\label{app:KL_modes}

\begin{figure}[h!]
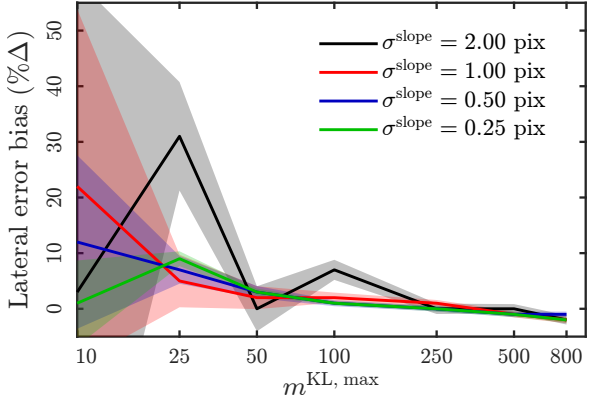
 % fig:2D_corr_sim_misreg_n_modes
        \centering
        
        % Internal command of the figure for the automatic sizing
	\newcommand{\PathFig}{Fig_2D_corr_sim_misreg_n_modes/}
    
    \newcommand{\subfigColor}{black}        
    
    \subfigimg[width=0.85\linewidth,pos=ul,font=\fontfig{\subfigColor}]{}{0.0}{\PathFig mod_sig_noise}
       
        \caption{\label{fig:2D_corr_sim_misreg_n_modes} Lateral error bias of the 2D correlation method in function of the noise $\sigma^\Tag{slope}$ on the slopes of the IM and the number of fitted modes. The coloured areas emphasise the $\pm\sigma$ regions.}
\end{figure}

Even if optimising the shape and the number of modes used in the modal fit was beyond the scope of the paper, the impact of the number of selected KL modes is presented in \reffig{fig:2D_corr_sim_misreg_n_modes} for different noise configurations $\sigma^\Tag{slope}$ on the IM slopes. For each case, 1000 random simulations were performed. To assess the noise propagation of the estimator, $\mUS$ was set to 100, leading to a theoretical resolution of one percent via the super-resolved map $\V{\alpha}$. The lateral shift was fixed to $\delta_{x} = \delta_{y} = \delta_{0} = 4.18\SApitch$. Thus the problem is symmetric and the figure presents the global statistics of  $\Brace{\tilde{\delta}_{x}-\delta_{0}, \tilde{\delta}_{y}-\delta_{0}}$.

The figure shows that the IM noise level has a limited impact. Indeed, for $m^{\Tag{KL, max}}\geq50$, the bias and the standard deviation of the estimator lie within the chosen target of~$1/8=\percent{12.5}$ of a subaperture for all the tested noise levels. The number of fitted modes appears as the key parameter of the method in order to limit the noise propagation and increase the achievable accuracy. This is expected since (1) increasing the number of modes allow us to average a larger number of noisy measurements, thus improving the S/N; (2) higher order modes show more high resolution features and thus gives access to a higher sensitivity. As a side remark, this figure also also supports the fact that any bias induced by the zero-padding strategy for the oversampling of the $\V{\alpha}$-map is marginal (a few percent) in regard to the initial resolution (the subaperture pitch~$\SApitch$).

\section{Cross-talk with other mis-registrations}
\label{app:cross_talk}

In this appendix, we introduce the impact of rotation and stretch misalignments on the performances of the proposed estimators. These two types of first order mis-registration are the most common ones with anamorphosis. This latter behaves as a stretch in our context and is consequently not presented.

\subsection{Perturbative 2D modal estimator}
\label{app:cross_talk_2D_corr}

The robustness of the 2D modal estimator in presence of other mis-registrations was assessed by simulating the convergence of a corrective loop on the lateral error. The initial lateral error was fixed to~$\Vdelta=\Paren{\delta_{x}, \delta_{y}}=\Paren{4.18, 3.73}\SApitch$. We kept the parameters of \refsec{sec:simu_2D_corr}: $\sigma^\Tag{slope}=0.25$ pixel, $m^{\Tag{KL, max}}=50$, $\mUS=8$. The corrective loop \refeq{eq:corrective_loop} is closed with a gain of~$\Thong = 1$.

\begin{figure}[h!]
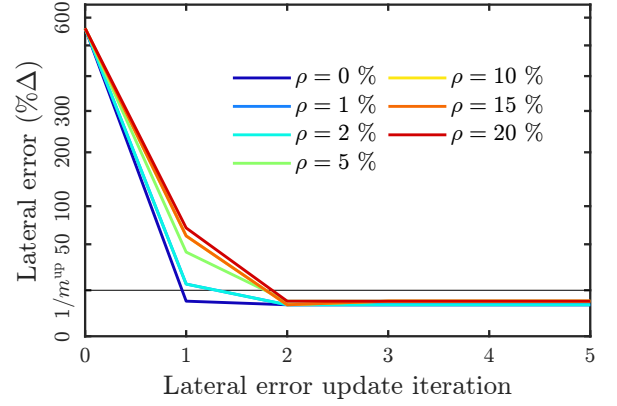
 % fig:2D_corr_sim_misreg_mag
        \centering
        
        % Internal command of the figure for the automatic sizing
	\newcommand{\PathFig}{Fig_2D_corr_sim_misreg_theta_mag/}
    
    \newcommand{\subfigColor}{black}        
    
    \subfigimg[width=0.85\linewidth,pos=ul,font=\fontfig{\subfigColor}]{}{0.0}{\PathFig Coeffs_sim_mag}
       
        \caption{\label{fig:2D_corr_sim_misreg_mag} Convergence of the lateral error after a few iterations of the corrective loop with the 2D modal estimator in function of the mis-registration magnification $\rho$ (square root scale).}
\end{figure}

\reffigsfull{fig:2D_corr_sim_misreg_mag}{fig:2D_corr_sim_misreg_theta} present the evolution of~$\Abs{\Vdelta}$ with the \mbox{iterations} of the corrective loop for different mis-registration magnifications and angles respectively. The estimator is robust to extreme stretches, with a convergence within the resolution of the estimator~$1/\mUS$ achieved in two iterations. Similar results are obtained on the angle up to a mis-registration of \SI{10}{\degree}. Performances then quickly degrade above this threshold.

\begin{figure}[h!]
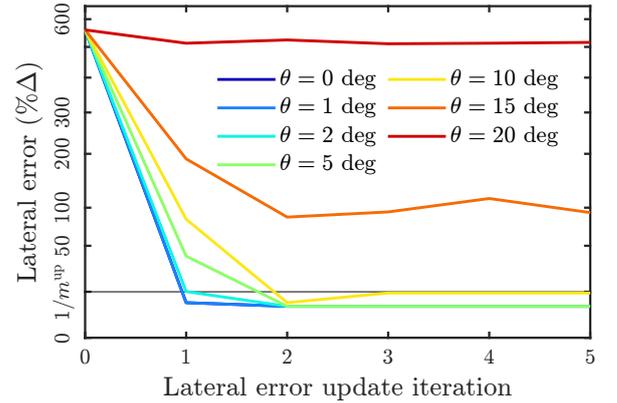
 % fig:2D_corr_sim_misreg_theta
        \centering
        
        % Internal command of the figure for the automatic sizing
	\newcommand{\PathFig}{Fig_2D_corr_sim_misreg_theta_mag/}
    
    \newcommand{\subfigColor}{black}        
    
    \subfigimg[width=0.85\linewidth,pos=ul,font=\fontfig{\subfigColor}]{}{0.0}{\PathFig Coeffs_sim_theta}
       
        \caption{\label{fig:2D_corr_sim_misreg_theta} {Convergence of the lateral error after a few iterations of the corrective loop with the 2D modal estimator in function of the mis-registration angle $\theta$ (square root scale).}}
\end{figure}

\subsection{Non-perturbative closed loop estimator}
\label{app:cross_talk_CL_corr}

For the closed loop estimator, the context is different than with the 2D modal estimator: any mis-registration must be small enough to permit the AO loop closure. Otherwise, the loop diverges and this estimator does not apply since such a situation generally triggers the loop opening by the system. In this context, any small mis-registration on the angle, $\delta_{\theta}$, or the stretch, $\delta_{\rho}$, is associated with the following transformation matrices:
\begin{equation}
        \label{eq:1st_misreg}
        \begin{bmatrix}
                1 & +\delta_{\theta}
                \\
                -\delta_{\theta} & 1
        \end{bmatrix}
        \text{ and }
        \begin{bmatrix}
                1+\delta_{\rho} & 0
                \\
                0 & 1+\delta_{\rho}
        \end{bmatrix}
        \,.
\end{equation}

\refeqfull{eq:1st_misreg} shows that a lateral error in a given direction will slightly leak (1) towards its perpendicular counterpart in presence of a rotation error and (2) towards its neighbouring spatial frequency in the same direction in presence of a magnification error. This leakage should be negligible, most of the correlation signal staying within its own spatial frequency.

Similar as for the 2D modal estimator, the robustness of the closed loop estimator was assessed by simulating the convergence of a corrective loop on the lateral error. The initial lateral error was fixed to~$\Vdelta=\Paren{\delta_{x}, \delta_{y}}=\percent{\Paren{25, 0}}\SApitch$. As in \refsec{sec:CL_estim_sim_wind}, batches of 500 closed loop iterations controlling 500 modes are gathered to feed the estimator and the corrective loop \refeq{eq:corrective_loop} is closed with a gain of~$\Thong = 0.5$.

\begin{figure}[h!]
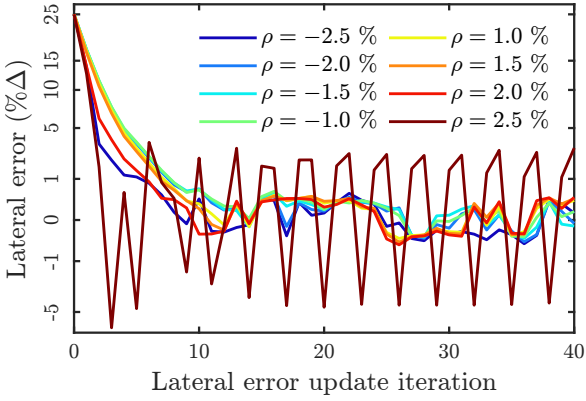
 % fig:2D_corr_sim_misreg_mag
        \centering
        
        % Internal command of the figure for the automatic sizing
	\newcommand{\PathFig}{Fig_CL_corr_misreg_theta_mag/}
    
    \newcommand{\subfigColor}{black}        
    
    \subfigimg[width=0.85\linewidth,pos=ul,font=\fontfig{\subfigColor}]{}{0.0}{\PathFig stretch_bias}
       
        \caption{\label{fig:CL_corr_misreg_mag} Convergence of the lateral error after a few iterations of the corrective loop with the closed loop estimator in function of the mis-registration magnification $\rho$ (square root scale).}
\end{figure}

\reffigsfull{fig:CL_corr_misreg_mag}{fig:CL_corr_misreg_theta} present the evolution of $\delta_{x}$ with the \mbox{iterations} of the corrective loop for different mis-registration magnifications and angles respectively. In all cases, the corrective loop successfully converges towards a null lateral error. When the loop is stable, for $\Abs{\theta}\leq\SI{1.5}{\degree}$ and $\Abs{\rho}\leq\percent{2}$, the accuracy is below $\percent{\pm 1}\SApitch$. Above these thresholds, the closed loop is unstable. Its divergence is prevented by the clipping of the DM commands. In this situation, the accuracy nevertheless stays \mbox{under} control, within $\percent{\pm 5}\SApitch$.

\begin{figure}[h!] % fig:2D_corr_sim_misreg_theta
        \centering
        
        % Internal command of the figure for the automatic sizing
	\newcommand{\PathFig}{Fig_CL_corr_misreg_theta_mag/}
    
    \newcommand{\subfigColor}{black}        
    
    \subfigimg[width=0.85\linewidth,pos=ul,font=\fontfig{\subfigColor}]{}{0.0}{\PathFig angle_bias}
       
        \caption{\label{fig:CL_corr_misreg_theta} Convergence of the lateral error after a few iterations of the corrective loop with the closed loop estimator in function of the mis-registration angle $\theta$ (square root scale).}
\end{figure}

\section{Closed loop transfer functions}
\label{app:transfer_functions}

In this appendix, we derive the transfer functions introduced in \refsec{sec:temp_coupling}. To ease the reading, we use the notation `$h\Paren{f}=\fdep{h}$'\footnote{In other words, $\fdep{h}$ is the temporal Fourier transform of $h\Paren{t}$ evaluated at the frequency $f$.} to highlight the dependency in terms of~$f$. As already discussed, we work under the assumptions of \refeq{eq:eq_12} that the transfer functions are identical for the two spatial modes. It is also implied that all this appendix is written for a given spatial frequency,~$\Vk$. 

In doing so, the diagram of \reffig{fig:CL_corr_spatial_coupling} is expressed as:
\begin{equation}
        \label{eq:spatial_coupling}
        \begin{cases}
                \begin{aligned}
                        \wf{1} = & -\ct\Paren{\fdep{A}\fdep{C}\Paren{\fdep{S}\wf{1}+\nf{1}}} + \pf{1}
                        \\
                        & - \st\Paren{\fdep{A}\fdep{C}\Paren{\fdep{S}\wf{2}+\nf{2}}}\,,
                \end{aligned}
                \\
                \begin{aligned}
                        \wf{2} = & -\ct\Paren{\fdep{A}\fdep{C}\Paren{\fdep{S}\wf{2}+\nf{2}}} + \pf{2}
                        \\
                        & + \st\Paren{\fdep{A}\fdep{C}\Paren{\fdep{S}\wf{1}+\nf{1}}}\,.
                \end{aligned}
        \end{cases}
\end{equation}
Introducing:
\begin{equation}
        \begin{cases}
                \muf \triangleq \fdep{A}\fdep{C}\fdep{S}
                \,,
                \\
                \Deltaf\Paren{\theta} \triangleq 1+2\muf\ct+\muf^{2}
                \,,  
        \end{cases}
\end{equation}
\refeq{eq:spatial_coupling} can be rewritten in a matrix shape:
\begin{equation}
        \V{M}\Paren{\theta}
        \begin{bmatrix}
                \wf{1}
                \\
                \wf{2}
        \end{bmatrix}
        =
        -\fdep{A}\fdep{C}\V{R}\Paren{\theta}
        \begin{bmatrix}
                \nf{1}
                \\
                \nf{2}
        \end{bmatrix}
        +
        \begin{bmatrix}
                \pf{1}
                \\
                \pf{2}
        \end{bmatrix}
        \,,
\end{equation}
with:
\begin{equation}
        \begin{cases}
                \fdep{\V{M}}\Paren{\theta}
                \triangleq &
                \begin{bmatrix}
                        1+\muf\ct & +\muf\st
                        \\
                        -\muf\st & 1+\muf\ct
                \end{bmatrix}
                \,,
                \\
                \V{R}\Paren{\theta}
                \triangleq &
                \begin{bmatrix}
                        +\ct & +\st
                        \\
                        -\st & +\ct
                \end{bmatrix}
                \,.
        \end{cases}
\end{equation}
Noticing that:
\begin{equation}
        \begin{cases}
                \fdep{\V{M}}\Inv\Paren{\theta}
                = & \Deltaf\Paren{\theta}\Inv\fdep{\V{M}}\Paren{-\theta}
                \,,
                \\
                \fdep{\V{M}}\Paren{-\theta}\V{R}\Paren{\theta}
                = &
                \begin{bmatrix}
                        \muf+\ct & +\st
                        \\
                        -\st & \muf+\ct
                \end{bmatrix}
                \,,
        \end{cases}
\end{equation}
it comes:
\begin{equation}
    \begin{aligned}
        \begin{bmatrix}
            \wf{1}
            \\
            \wf{2}
        \end{bmatrix}
        = &
        -\frac{\fdep{A}\fdep{C}}{\Deltaf\Paren{\theta}}
        \begin{bmatrix}
            \muf+\ct & +\st
            \\
            -\st & \muf+\ct
        \end{bmatrix}
        \begin{bmatrix}
            \nf{1}
            \\
            \nf{2}
        \end{bmatrix}
        \\
        & + \Deltaf\Paren{\theta}\Inv\fdep{\V{M}}\Paren{-\theta}
        \begin{bmatrix}
            \pf{1}
            \\
            \pf{2}
        \end{bmatrix}  
        \,,
    \end{aligned}
\end{equation}
and from $\mf{i} = \fdep{S}\wf{i}+\nf{i}$ for~$i\in\Brace{1,2}$, we finally get:
\begin{equation}
    \label{eq:transfer_functions}
    \begin{aligned}
        \begin{bmatrix}
            \mf{1}
            \\
            \mf{2}
        \end{bmatrix}
        = &
        \begin{bmatrix}
            \nf{1}
            \\
            \nf{2}
        \end{bmatrix}
        -\frac{\muf}{\Deltaf\Paren{\theta}}
        \begin{bmatrix}
            \muf+\ct & +\st
            \\
            -\st & \muf+\ct
        \end{bmatrix}
        \begin{bmatrix}
            \nf{1}
            \\
            \nf{2}
        \end{bmatrix}
        \\
        & + \frac{\fdep{S}}{\Deltaf\Paren{\theta}}
        \begin{bmatrix}
            1+\muf\ct & -\muf\st
            \\
            +\muf\st & 1+\muf\ct
        \end{bmatrix}
        \begin{bmatrix}
            \pf{1}
            \\
            \pf{2}
        \end{bmatrix}
       \,.
    \end{aligned}
\end{equation}
As the perturbations~$\pf{i}$ and noises~$\nf{j}$ are independent, the crossproduct of two different terms is null:
\begin{equation}
    \Avg{\pf{i}\nfconj{j}}=\Avg{\pf{i}}\Avg{\nfconj{j}}=0
    \,.
\end{equation}
Similarly, the noise in the two parts of the diagram of \reffig{fig:CL_corr_spatial_coupling} are independent and:
\begin{equation}
    \Avg{\nf{1}\nfconj{2}}=0
    \,.
\end{equation}
As discussed below, only~$\Avg{\pf{1}\pfconj{2}}$ may be different from zero. Finally, one can assume that the pairs of noise terms~$\Paren{\nf{1},\nf{2}}$ and~$\Paren{\pf{1},\pf{2}}$ have identical properties, leading to:
\begin{equation}
    \label{eq:eq_noise}
    \begin{cases}
        \Avg{\nf{1}\nfconj{1}}=\Avg{\nf{2}\nfconj{2}}=\Vvar{\nf{}}
        \,,
        \\
        \Avg{\pf{1}\pfconj{1}}=\Avg{\pf{2}\pfconj{2}}=\Vvar{\pf{}}
        \,.
    \end{cases}
\end{equation}
It is now possible to compute the expected covariances $\Avg{\mf{i}\mfconj{j}}$ for different regimes.

\subsection{Noise propagation}

In the noise-limited regime, the turbulence perturbation terms~$\pf{i}$ are assumed to be negligible.  It comes from \refeqs{eq:transfer_functions}{eq:eq_noise} that:
\begin{equation}
    \begin{cases}
        \Avg{\mf{i}\mfconj{i}} = \Paren{\Abs{\frac{1+\muf\ct}{\Deltaf\Paren{\theta}}}^2 + \Abs{\frac{\muf}{\Deltaf\Paren{\theta}}\st}^2}\Vvar{\nf{}}
        \,,
        \\
        \Avg{\mf{1}\mfconj{2}} = 2i\imag{\frac{1+\muf\ct}{\Abs{\Deltaf\Paren{\theta}}^2}\mufconj\st}\Vvar{\nf{}}
        \,.
    \end{cases}
\end{equation}
Then, using \refeqs{eq:eq_12}{eq:corr}, the correlation between~$\cf{1}$ and~$\cf{2}$ is written as:
\begin{align}
    \Vcorr{\cf{1},\cf{2}}\Paren{\theta}
    & {}={} \Vcorr{\mf{1},\mf{2}}\Paren{\theta} 
    \\
    \label{eq:corr_theta}
    & {}={} 2i\st\frac{\imag{\Paren{1+\muf\ct}\mufconj}}{\Abs{1+\muf\ct}^2 + \Abs{\muf\st}^2}
    \,.
\end{align}
And finally, for a mis-registration parameter tending towards zero:
\begin{equation}
    \label{eq:corr_00}
    \Vcorr{\cf{1},\cf{2}}\Paren{\theta} / \theta \underset{\theta\rightarrow0}{\sim} 2i\imag{\frac{\mufconj}{1+\mufconj}} = i \Vcorr{0}\Paren{f}
    \,.
\end{equation}

\subsection{Turbulence propagation}
\label{app:turb_prop}

In the situation where the system noise is dominated by the turbulence disturbance, $\nf{i}$ are assumed to be negligible. It comes from \refeqs{eq:transfer_functions}{eq:eq_noise}:
\begin{equation}
    \label{eq:corr_wind}
    \begin{cases}
        \begin{aligned}
            \Avg{\mf{1}\mfconj{1}} {}={} & \Abs{\frac{\fdep{S}}{\Deltaf\Paren{\theta}}}^2 \bigg[ \Paren{\Abs{1+\muf\ct}^2 + \Abs{\muf\st}^2}\Vvar{\pf{}}
            \\
            & - 2\real{\Paren{1+\muf\ct}\mufconj\st\Avg{\pf{1}\pfconj{2}}}\bigg]
            \,,
        \end{aligned}
        \\
        \begin{aligned}
            \Avg{\mf{2}\mfconj{2}} {}={} & \Abs{\frac{\fdep{S}}{\Deltaf\Paren{\theta}}}^2 \bigg[\Paren{\Abs{1+\muf\ct}^2 + \Abs{\muf\st}^2}\Vvar{\pf{}}
            \\
            & + 2
            \real{\Paren{1+\mufconj\cos\theta}\muf\sin\theta\Avg{\pf{1}\pfconj{2}}}\bigg]
            \,,
        \end{aligned}
        \\
        \begin{aligned}
            \Avg{\mf{1}\mfconj{2}} {}={} & \Abs{\frac{\fdep{S}}{\Deltaf\Paren{\theta}}}^2 \bigg[
            2i\imag{\Paren{1+\muf\ct}\mufconj\st}\Vvar{\pf{}}
            \\
            & + \Abs{1+\muf\ct}^2\Avg{\pf{1}\pfconj{2}} 
            \\
            & - \Abs{\muf\st}^2\Avg{\pf{2}\pfconj{1}}\bigg]
            \,.
        \end{aligned}
    \end{cases}
    \,
\end{equation}

In the situation where the turbulence perturbations are independent, $\Avg{\pf{1}\pfconj{2}} = 0$, \refeq{eq:corr_wind} leads to the same results than for the noise limited regime of the previous section:
\begin{equation}
    \begin{cases}
        \Vcorr{\cf{1},\cf{2}}\Paren{\theta} =  \Vcorr{\mf{1},\mf{2}}\Paren{\theta} = 2i\st\frac{\imag{\Paren{1+\muf\ct}\mufconj}}{\Abs{1+\muf\ct}^2 + \Abs{\muf\st}^2}
        \,,
        \\
        \Vcorr{\cf{1},\cf{2}}\Paren{\theta} / \theta \underset{\theta\rightarrow0}{\sim} 2i\imag{\frac{\mufconj}{1+\mufconj}}
         = i \Vcorr{0}\Paren{f}
         \,.
    \end{cases}
    \,
\end{equation}
This result was expected since in these two situations, $\nf{i}$ and $\pf{i}$ play a similar role of pure independent noises injected in the system with just a normalisation by $\fdep{S}$.  This normalisation factor is cancelled out when estimating the correlation via \refeq{eq:corr}.

Nonetheless, $\pf{1}$ and $\pf{2}$ are not always independent. This is the case for a pure frozen flow, the typical worst case offender of mis-registration estimators \citep[Sect. 4.3]{Heritier:19_frozen_wind, Heritier:19_PHD}. Indeed, in such a situation, the wavefront is expressed as:
\begin{align}
    \Vw {}={} & w_0\cos\Paren{2\pi\Vk\scaprod\Paren{\Vx-\Vv t}-\varphi_0}
    \\
    {}={} & w_0\cos\Paren{2\pi \fwind t + \varphi_0}\Vw_{1} + w_0\sin\Paren{2\pi \fwind t + \varphi_0}\Vw_{2}
    \,,
\end{align}
with~$\varphi_0$ a constant phase delay and with $\fwind$ the temporal frequency of the wind projected parallel to the spatial frequency,~$\Vk$:
\begin{equation}
    \fwind = \Vk\scaprod\Vv
    \,.
\end{equation}
Thus, 
\begin{equation}
    \begin{cases}
       p_1\Paren{t} = w_0\cos\Paren{2\pi \fwind t + \varphi_0}
       \,,
        \\
       p_2\Paren{t} = w_0\sin\Paren{2\pi \fwind t + \varphi_0}
       \,,
    \end{cases}
\end{equation}
which, after a Fourier transform, gives in the frequency space:
\begin{equation}
    \begin{cases}
       p_1\Paren{f} = \frac{w_0}{2} \Paren{e^{i\varphi_0} \dirfunc{\fwind}\Paren{f} + e^{-i\varphi_0} \dirfunc{-\fwind}\Paren{f}}
       \,,
        \\
       p_2\Paren{f} = \frac{w_0}{2i} \Paren{e^{i\varphi_0} \dirfunc{\fwind}\Paren{f} - e^{-i\varphi_0} \dirfunc{-\fwind}\Paren{f}}
       \,,
    \end{cases}
\end{equation}
where~$\dirfunc{x}$ is the Dirac function centered in~$x$. Thus, we finally have:
\begin{equation}
    \begin{cases}
        \forall \Abs{f} \neq \fwind, & \Avg{p_{1}\conj{p}_2}\Paren{f} = 0
        \,,
        \\
        f=\fwind, & \corr{c_1,c_2}\Paren{\fwind} \underset{\theta\rightarrow0}{\sim} i
        \,,
        \\
        f=-\fwind, & \corr{c_1,c_2}\Paren{-\fwind} \underset{\theta\rightarrow0}{\sim} -i
        \,.
    \end{cases}
\end{equation}
This means that a frozen flow turbulence has a minimal impact on the empirical correlation of \refeq{eq:corr_empirical}. Only frequencies close to $f\simeq\Vk\scaprod\Vv$ should be impacted. In addition, these frequencies depend on the considered spatial mode~$\Vk$ and consequently the impact on the estimator of \refeq{eq:CL_estim} should be diluted among the controlled frequencies~$\Vk\in\kctrl$.

\begin{figure}[t!] % fig:CL_corr_wind_corr
    \centering
    
     % Internal command of the figure for the automatic sizing
    % Path of the files
    \newcommand{\PathFig}{Fig_CL_corr_wind_corr/}
    
    \newcommand{\subfigColor}{black}
    
    % Line ratio
    \newcommand{\LineRatio}{0.985}
    
    \newcommand{\figOne}{\PathFig corr_curv}
    \newcommand{\figTwo}{\PathFig corr_map}
    
    % Getting the size of the boxes
    \sbox1{\includegraphics{\figOne}}
    \sbox2{\includegraphics{\figTwo}}
    
    % Defining column width command
    \newcommand{\ColumnWidth}[1]
        {\dimexpr \LineRatio \linewidth * \AspectRatio{#1} / (\AspectRatio{1} + \AspectRatio{2}) \relax
        }
    \newcommand{\ColumnGap}{\hspace {\dimexpr \linewidth /1 - \LineRatio\linewidth /1 }}

    % Figure table
    \begin{tabular}{
        @{}
        M{\ColumnWidth{1}}
        @{\ColumnGap}
        M{\ColumnWidth{2}}
        @{}
        }
        
        \subfigimg[width=\linewidth,pos=bl,font=\fontfig{\subfigColor}]{}{0.0}{\figOne} &
        \subfigimg[width=\linewidth,pos=ul,font=\fontfig{\subfigColor}]{}{0.0}{\figTwo}
        \\[-0.15cm]
    \end{tabular}
    \begin{tabular}{
        @{}
        M{\ColumnWidth{1}}
        @{\ColumnGap}
        M{\ColumnWidth{2}}
        @{}
        }
        $\;\;\;$Frequency $f$ (\SI{}{\hertz})
        &
        Spatial frequency $\Vk$ %(\SI{}{\per\meter})
    \end{tabular} 
          
    \caption{\label{fig:CL_corr_wind_corr} Impact on the temporal correlation curves $\Vcorrt{t}$ (left panel) and the map of the best fit correlation coefficients $\Vcorrt{2D}$ (right panel) of a frozen flow perturbation of wind speed~$\vWind=\SI{15}{\meter\per\second}$ aligned with a lateral error of $\delta_x=\percent{-20}\SApitch$ in a noise limited regime $\nph=10$ and controlling~$\nmodes=800$. Coloured curves: best temporal correlation fit~$\Vcorrt{t}$ for different sets of controlled modes (mean filter with a sliding window of $\pm3$ data points). Black curves: theoretical correlation curve~$\Vcorr{0}$. The coloured dashed lines emphasise the temporal frequencies~$\fwind = k_0\vWind$ associated with the spatial frequencies~$k_0$ highlighted on the right panel.
    }
\end{figure}

These different features are illustrated in \reffig{fig:CL_corr_wind_corr} for a noise limited regime $\nph=10$ and controlling~$\nmodes=800$. The frozen turbulence flows at a speed of~$\vWind=\SI{15}{\meter\per\second}$ in the same direction than a lateral error of~$\delta_x=\percent{-20}\SApitch$. It seems that the wind has no qualitative impact on the best fit correlation map~$\Vcorrt{2D}$ (right panel of the figure). The purple temporal correlation curve~$\Vcorrt{t}$ is given by \refeq{eq:corr_t}, while the other temporal correlation curves are obtained by replacing~$\Vk\in\kctrl$ in the equation by:
\begin{equation}
    \Brace{\Vk \text{ s.t. } \Vk\in\kctrl \text{ and } k_x = k_0}
    \,.
\end{equation}
Different values of~$k_0$ are displayed: $5D^{-1}$ (red),  $10 D^{-1}$ (green),  $15 D^{-1}$ (blue). As expected, the frozen turbulence has only a limited impact on the empirical correlation which is centered around~$\fwind = k_0\vWind$. The rest of the curve still matches  the theoretical asymptote of~$\Vcorr{0}$ well. The same conclusion applies to the global fit, in purple. Overall, only the low temporal frequencies are impacted.

Finally, the estimated lateral error is~$\tilde{\Vdelta}=\percent{\Paren{-11.5,0}}\SApitch$. This is in the same order of magnitude than the results of \reffig{fig:CL_corr_gain}. This confirms that a wind blowing parallel to the lateral error should have a limited impact on the proposed closed loop estimator.

\section{Applying photon noise onto the SH-WFS slopes}
\label{app:slope_noise}

In this appendix, we describe our method to add the photon noise directly onto the slopes of the geometrical model of the SH-WFS. For a given number of photon~$\nph$, the expected standard deviation of the wavefront phase on a subaperture is~\citep{Rousset:99_ph_noise}:\begin{equation}
        \sigma^\Tag{wf} = \frac{\pi}{\sqrt{2\nph}}\frac{\Theta^\Tag{spot}}{\Theta^\Tag{dif}} \quad \Paren{\text{phase radian}}
        \,,
\end{equation}
where $\Theta^\Tag{spot}$ and $\Theta^\Tag{dif}$ are the full widths at half maximum of the spot and the diffraction limited patterns of a subaperture. Noting~$d^\Tag{sub}$ the diameter of a subaperture, it becomes:
\begin{equation}
        \begin{cases}
        \Theta^\Tag{spot} \simeq \frac{\lambda^\Tag{wfs}}{\rFried\Paren{\lambda^\Tag{wfs}}}
        \,,
        \\
        \Theta^\Tag{dif} \simeq \frac{\lambda^\Tag{wfs}}{d^\Tag{sub}}
        \,,
        \end{cases}
\end{equation}
where~$\rFried\Paren{\lambda^\Tag{wfs}}$ is the Fried parameter scaled to the WFS wavelength.
This leads to an optical path difference of: 
\begin{equation}
        \sigma^\Tag{OPD} = \frac{\lambda^\Tag{wfs}}{2}\frac{d^\Tag{sub}}{\rFried\Paren{\lambda^\Tag{wfs}}}\frac{1}{\sqrt{2\nph}}
        \,,
\end{equation}
and thus an angle across the subaperture diameter of:
\begin{equation}
        \label{eq:sig_ph}
        \sigma^\Tag{ph} = \frac{\lambda^\Tag{wfs}}{2\rFried\Paren{\lambda^\Tag{wfs}}}\frac{1}{\sqrt{2\nph}} \quad \Paren{\text{angle radian}}.
\end{equation}
The different equivalent photon noises induced on the slopes by the number of photons are given in \reftab{tab:sig_ph}.

\begin{table}[h!] % tab:sig_ph
        \caption{\label{tab:sig_ph} Photon noise on the slopes for the different simulated numbers of photons.}
        \centering
        \begin{tabular}{cc}
        \hline
        \hline
        $\nph$ & $\sigma^\Tag{ph}\Paren{\text{mas}}$ 
    \\
        \hline
        10 & 88.6 
        \\
        100 & 28.0
        \\
        1000 & 8.86 
        \\
        \hline
        \end{tabular}
        %\tablefoot{Coucou}
\end{table}

\section{Impact of the telemetry batch size on the closed loop estimator}
\label{app:batch_size}

The influence of the telemetry batch size on the closed loop estimator was assessed in simulation by introducing a fixed lateral error of~$\Vdelta=\Paren{\delta_{x}, \delta_{y}}=\percent{\Paren{10, 0}}\SApitch$. For each tested batch size, 100 batches in a pure noise regime were gathered, controlling 500 modes.

Their statistics are shown in \reftab{tab:batch_size}. First, the sensitivity of~$\gcl \simeq 0.7$ of the estimator can be seen with $\Paren{\tilde{\delta}_{x}, \tilde{\delta}_{y}}\simeq \percent{\Paren{6.9, 0.0}}\SApitch$, in agreement with \reffig{fig:CL_corr_gain}. Then, as expected, the noise on the estimated value decrease with the number of frames in the batch and is consistent on the $x$ and $y$-axes. The size of 500 frames, used in this work, gives a standard deviation of $\pm \percent{0.25}\SApitch$. This was considered to be a good compromise between speed and accuracy, allowing for the corrective loop on the lateral error to be estimated and updated once every second.

\begin{table}[t!] % tab:batch_size
        \caption{\label{tab:batch_size} Statistics of the estimated lateral error in function of the telemetry batch size.}
        \centering
        \begin{tabular}{ccc}
        \hline
        \hline
    Batch size      & $\tilde{\delta}_x$ $\Paren{\%\SApitch}$& $\tilde{\delta}_y$ $\Paren{\%\SApitch}$\\
        \hline
   $  50$ frames    &  $6.49 \pm 0.89$  &  $-0.06 \pm 0.83$ \\
   $ 100$ frames    &  $6.68 \pm 0.55$  &  $+0.01 \pm 0.56$ \\
   $ 200$ frames    &  $6.79 \pm 0.37$  &  $-0.03 \pm 0.39$ \\
   $ 500$ frames    &  $6.90 \pm 0.25$  &  $-0.02 \pm 0.23$ \\
   $1000$ frames    &  $6.88 \pm 0.18$  &  $-0.02 \pm 0.16$ \\
   $2000$ frames    &  $6.91 \pm 0.13$  &  $-0.01 \pm 0.13$ \\
   $5000$ frames    &  $6.91 \pm 0.09$  &  $-0.00 \pm 0.08$ \\
        \hline
      Truth         &  $\delta_x = 10$  &  $\delta_y =  0$ \\ 
        \hline
        \end{tabular}
        %\tablefoot{Coucou}
\end{table}

\begin{figure}[t!]
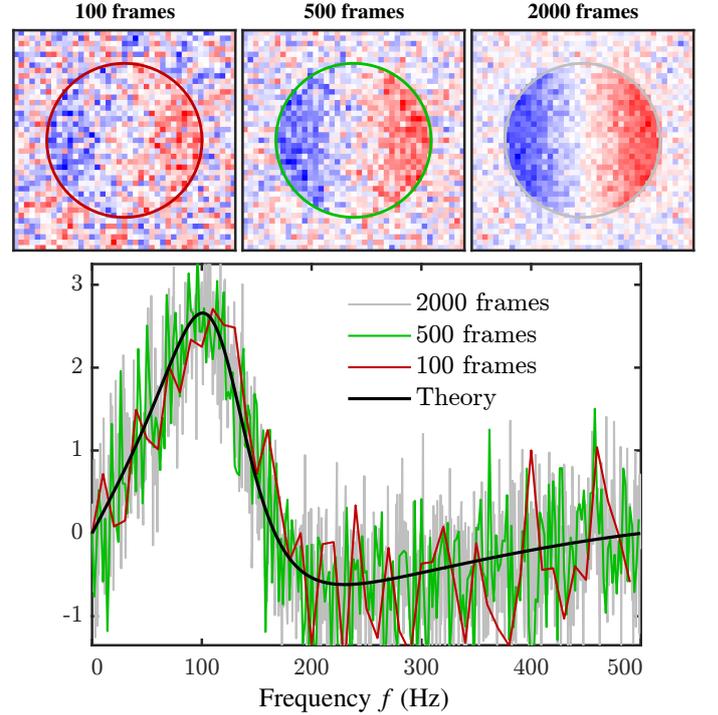
 % fig:batch_size
    \centering
    
    % Internal command of the figure for the automatic sizing
    % Path of the files
    \newcommand{\PathFig}{Fig_CL_corr_misreg_nb_loop/}

    % Vertical space between lines
    \newcommand{\spaceLine}{-0.05cm}
    
    % Font of the text in the figure
    \newcommand{\fontTxt}[1]{\textbf{\tiny #1}}
    
    % Line ratio
    \newcommand{\LineRatio}{0.99}
    % Color of the panel letter
    \newcommand{\subfigColor}{black}
    % Defining column width command
    \newcommand{\ColumnWidth}
        {\dimexpr \LineRatio \linewidth /3 \relax
        }
    \newcommand{\ColumnGap}{\hspace {\dimexpr \linewidth /2 - \LineRatio\linewidth /2 }}
    % Figure table
    \begin{tabular}{
        @{}
        M{\ColumnWidth}
        @{\ColumnGap}
        M{\ColumnWidth}
        @{\ColumnGap}
        M{\ColumnWidth}
        @{}
        }
        
        % Title line    
        \fontTxt{100 frames}
        &
        \fontTxt{500 frames}
        &
        \fontTxt{2000 frames}
        \\             
        \subfigimg[width=\linewidth,pos=ul,font=\fontfig{\subfigColor}]{$\,$}{0.0}{\PathFig corr_map_100_frames} &
        \subfigimg[width=\linewidth,pos=ul,font=\fontfig{\subfigColor}]{$\,$}{0.0}{\PathFig corr_map_500_frames} &
        \subfigimg[width=\linewidth,pos=ul,font=\fontfig{\subfigColor}]{$\,$}{0.0}{\PathFig corr_map_2000_frames}
    \end{tabular}
    
    \vspace{\spaceLine}
	\subfigimg[width=0.85\linewidth,pos=ul,font=\fontfig{\subfigColor}]{$\quad\;\;$}{0.0}{\PathFig corr_curv}
	$\;\;\;$Frequency $f$ (\SI{}{\hertz})
    
    \caption{\label{fig:batch_size} Impact of the telemetry batch size on the map of the best fit correlation coefficients $\Vcorrt{2D}$ (top panels) and the temporal correlation curves $\Vcorrt{t}$ (bottom panel). The coloured circles emphasise the controlled frequency space.}
\end{figure}

For information, \reffig{fig:batch_size} presents an example of the correlation patterns obtained when varying the size of the telemetry batch. The increase of the signal over noise ratio and the frequency resolution with the number of frames can clearly be seen in the correlation maps~$\Vcorrt{2D}$  (top panels) and temporal correlation curves~$\Vcorrt{t}$ (bottom panel).

\section{Estimating the lateral bias induced by the wind}
\label{app:wind_conv}

In this appendix, we detail how the convergence values of the bias induced by the frozen flow wind are obtained from the iterative updates of the lateral errors with time. For~$\nph=10$ and 100, $n^\Tag{it}=40$ updates (spanning over \SI{20}{\second}) of~$\Vdelta$ were performed. For~$\nph=1000$, the convergence is slower and $n^\Tag{it}=60$ updates (spanning over \SI{30}{\second}) of~$\Vdelta$ were performed.

\begin{figure}[th!]
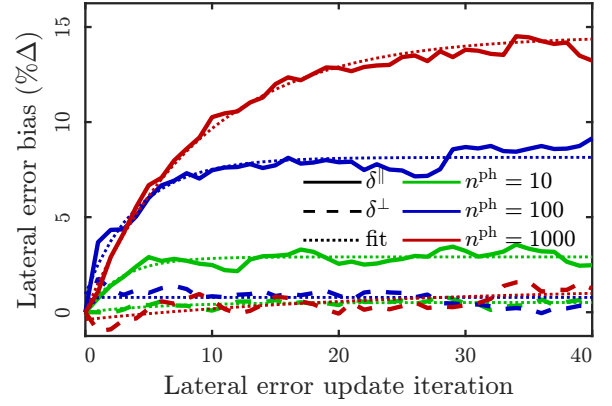
 % fig:CL_corr_wind_bias_conv
        \centering
        
        % Internal command of the figure for the automatic sizing
        % Path of the files
        \newcommand{\PathFig}{Fig_CL_corr_wind_bias/}
        
	\newcommand{\subfigColor}{black}        
        
        \subfigimg[width=0.85\linewidth,pos=ul,font=\fontfig{\subfigColor}]{$\quad\;\;$}{0.0}{\PathFig wind_conv_v_20_ang_30}
        \caption{\label{fig:CL_corr_wind_bias_conv} Convergence of the lateral error~$\Vdelta$ in percent of subaperture pitch~$\SApitch$ in function of the update iteration for $\vWind=\SI{20}{\meter\per\second}$, $\thetaWind=\SI{30}{\degree}$, and for different photon noise levels~$\nph$ (colours). Plain (resp. dashed) curves: lateral error~$\delta^\parallel$ parallel (resp. ~$\delta^\perp$ perpendicular) to the wind direction. Dotted curves: fitted exponential law.}
\end{figure}

\reffigfull{fig:CL_corr_wind_bias_conv} shows the evolution of~$\Vdelta$ with the update iterations for different level of photon noise. Convergence is achieved despite being noisy. To robustly estimate the convergence value, an exponential law (dotted curves in the figure) is fitted in the data points of each direction parallel $\parallel$ and perpendicular $\perp$ to the wind's direction:
\begin{equation}
        \delta^{i}_{\parallel, \perp} \simeq \Nguu_{\parallel, \perp} \Paren{1-e^{-\Reua_{\parallel, \perp}\Paren{i-i^0_{\parallel, \perp}}}}
        \,.
\end{equation}
The fit of~$\Paren{\Nguu, \Reua, i^0}$ is performed by minimising the root mean square difference between the model and the simulation with the simplex search method of \citet{Lagarias:98_fminsearch}, enforcing that~$1/\Reua < 25$. \Nguu{} gives the convergence value and the bias induced by the wind on the closed loop estimator.

\end{appendix}

\end{document}